\documentclass[12pt,preprint]{aastex}

\newcommand{\ssize}[1][opt]{\scriptsize}
\newcommand{\tnm}[1][]{\tablenotemark}
\newcommand{\tnt}[1][]{\tablenotetext}
\newcommand{\ph}[1][]{\phantom}
\newcommand{\fns}[1][opt]{\footnotesize}

\begin{document}

\title{H$_{\mathbf{2}}$O MASER OBSERVATIONS OF CANDIDATE POST-AGB STARS \\
AND\\
DISCOVERY OF THREE HIGH-VELOCITY WATER SOURCES}

\shorttitle{H$_{\mathbf{2}}$O MASER OBSERVATIONS OF POST-AGB STARS}
\shortauthors{Deacon et. al.}

\author{R. M. Deacon\altaffilmark{1}}
\affil{School of Physics A29, University of Sydney, NSW 2006, Australia}
\altaffiltext{1}{Affiliated with the CSIRO Australia Telescope National Facility}
\email{R.Deacon@physics.usyd.edu.au}
\author{J. M. Chapman}
\affil{CSIRO Australia Telescope National Facility, P.O. Box 76, Epping, NSW 1710, Australia}
\author{A. J. Green}
\affil{School of Physics A29, University of Sydney, NSW 2006, Australia}
\author{M. N. Sevenster}
\affil{Sterrewacht Leiden, Niels Bohrweg 2, 2333 RA Leiden, The Netherlands}

\begin{abstract}
We present the results of 22~GHz H$_2$O maser observations of a sample
of 85 post-Asymptotic Giant Branch (post-AGB) candidate stars,
selected on the basis of their OH 1612 MHz maser and far-infrared
properties. All sources were observed with the Tidbinbilla 70-m radio
telescope and 21 detections were made. 86~GHz SiO Mopra observations
of a subset of the sample are also presented. Of the 21 H$_2$O
detections, 15 are from sources that are likely to be massive AGB
stars and most of these show typical, regular H$_2$O maser
profiles. In contrast, nearly all the detections of more evolved stars
exhibited high-velocity H$_2$O maser emission. Of the five sources
seen, v223 (W43A, IRAS 18450$-$0148) is a well known `water-fountain'
source which belongs to a small group of post-AGB stars with highly
collimated, high-velocity H$_2$O maser emission. A second source in
our sample, v270 (IRAS 18596$+$0315), is also known to have
high-velocity emission. We report the discovery of similar emission
from a further three sources, d46 (IRAS 15445$-$5449), d62 (IRAS
15544$-$5332) and b292 (IRAS 18043$-$2116). The source d46 is an
evolved post-AGB star with highly unusual maser properties. The H$_2$O
maser emission from d62 is probably associated with a massive star.
The source b292 is a young post-AGB star that is highly likely to be a
water-fountain source, with masers detected over a velocity range of
210 km s$^{-1}$.
\end{abstract}

\keywords{masers --- stars: AGB and post-AGB --- stars: late-type --- 
stars: winds, outflows --- planetary nebulae: general --- 
radio lines: stars}

\section{INTRODUCTION} \label{introduction}

Planetary nebulae (PN) are seen as the clouds of ionised gas that
surround the central cores of dying stars.  The nebulae often show
beautiful and complex structures: while some appear circular, more
than half have elliptical or bipolar shapes, in some cases with
complex, filamentary structures \citep{Man00,Kwo00}.  The cause of the
diverse range of PN morphologies is still not well understood but is
likely to be associated with rapid changes that occur as a star
evolves from the Asymptotic Giant Branch (AGB) to the PN stage.

Most stars with an initial Main Sequence mass below about eight solar
masses (M$_{\odot}$) will evolve to become PN.  Towards the end of the
AGB stage of stellar evolution, stars lose a significant fraction of
their initial mass through pulsation-driven mass loss \citep{Vas93}.
The Mira variables are optically-visible long-period variable AGB
stars with pulsation periods of $\sim$ 200 -- 800 days, mass-loss
rates of $\sim$ 10$^{-8}$ -- 10$^{-6}$ M$_{\odot}$ yr$^{-1}$ and
envelope expansion velocities of 5 -- 10 km s$^{-1}$. For AGB stars
with higher mass-loss rates, the central stars are generally invisible
but their circumstellar envelopes can be detected through their
infrared and/or radio maser emission \citep{Cha95}. The OH/IR stars
are AGB stars that extend the properties of Miras with thicker
circumstellar envelopes, higher mass-loss rates (10$^{-7}$ --
10$^{-4}$ M$_{\odot}$ yr$^{-1}$), longer pulsation periods (300 --
3000 days) and higher expansion velocities (5 -- 30 km s$^{-1}$). More
than 1500 OH/IR stars have been detected from OH maser surveys while
far-infrared emission from many thousands of AGB stars has been
detected in the InfraRed Astronomical Satellite (IRAS) and Midcourse
Space eXperiment (MSX) surveys \citep{Van88,Hab96}.

In the oxygen-rich circumstellar envelopes of Miras and OH/IR stars,
OH, H$_2$O and SiO maser emission may be detected. The SiO maser
emission originates in the upper stellar atmospheres at heights of
several stellar radii while the H$_2$O maser emission occurs further
out, from the warm inner regions of the circumstellar envelopes.  The
OH 1612 MHz masers occur at even greater distances, in the cooler
outer envelopes at typically several hundred stellar radii
\citep{Hab96}.  The OH 1612 MHz spectra of OH/IR stars almost
invariably show a double-peaked profile consistent with a spherical
outflow.  For example, 86 per cent of 766 sources detected in an OH
1612 MHz survey of the Galactic Plane have the canonical double-peaked
profiles \citep{Sev97a,Sev97b,Sev01a}.

The end of the AGB evolutionary stage probably occurs when so much
mass has been lost that a star can no longer support radial
pulsations.  The mass-loss rate then greatly decreases and the surface
temperature of the star rises rapidly.  During this `post-AGB' phase,
the star changes from losing mass in a slow, dense wind, with a
velocity of typically 15 km s$^{-1}$, to mass loss in a hot, low
density wind with a velocity of several hundred km~s$^{-1}$.
The hot wind sweeps up and compresses the remnant circumstellar
envelope while the central star may again become optically visible as
radiation from the star ionises part or all of the remnant envelope
\citep[e.g.][]{Kwo78}.  These older post-AGB objects that have become 
optically visible but are not yet fully-ionised PN are often referred
to as `proto-planetary nebula' (PPN).

Several mechanisms have been used to explain the production of
non-spherical PN.  If the slow AGB wind possesses a weak asymmetry
with a dense equatorial region, then a bipolar structure may be
produced and amplified during the post-AGB phase by interaction with a
fast wind \citep[e.g.][]{Fra94b}.  Magnetic fields from the stars may
also constrain or collimate the stellar winds \citep[e.g.][]{Gar99}.
In some cases, bipolarity may also occur due to the presence of
companion stars or planets \citep{Woo00,Sok01}.

The hydrodynamical models for interacting stellar winds (ISW) do not
predict that strong bipolar structures are formed in the early
post-AGB phase, as the central stars are still too cool ($<$~15 000 K)
to produce the fast wind.  However, \citet{Sah98} suggested that fast
jets {\it can} operate in the early post-AGB stage and are the
dominant shaping mechanism.  The origin of these jets is unclear,
although in recent years there has been increasing evidence for high
velocity, short-lived jets from post-AGB stars.  In some cases the
jets are traced by OH or H$_2$O maser emission or through radio
continuum \citep[e.g.][]{Mir01,Ima02,Ima04,Bob05}.

Maser observations provide a powerful tool for identifying and
investigating the onset of wind asymmetries during the early post-AGB
evolutionary stage.  However, to date there have been few systematic
studies of the maser properties of post-AGB stars as most studies have
concentrated on sources selected to have unusual OH maser properties.
We are studying the maser properties of a well-defined sample of 85
candidate post-AGB stars.  The sample was selected from sources likely to 
be post-AGB stars, based on their far-infrared IRAS and MSX properties
\citep{Sev02a,Sev02b}, which were detected in an OH 1612 MHz survey of the 
Galactic Plane with the Australia Telescope Compact Array and Very Large
Telescope \citep[hereafter the ATCA/VLA survey,][]{Sev97a,Sev97b,Sev01a}.

In \citet[][hereafter Paper I]{Dea04}, the results from single-dish OH
1612, 1665, 1667 and 1720 MHz maser observations were presented.  All
sources were detected again at 1612 MHz, while 27 sources were measured at
1665 MHz and 47 at 1667~MHz.  Twenty five per cent of sources in this
sample have maser profiles indicative of aspherical wind morphologies.
A possible trend was identified in which some sources with classic peaked
OH 1612~MHz spectral profiles  evolve into profiles characteristic of 
developed bipolar outflows.

In this paper, results from single-dish 22 GHz H$_2$O maser
observations for the complete sample are presented.  A subset of 11
sources has also been observed at the SiO maser line near 86 GHz and
the detections are also reported.  The results from a previous radio
continuum detection of one source (d46) in 1998 are also discussed.

\subsection{H$_2$O Maser Emission}
 
The 22 GHz maser line comes from the $6_{16}$--$5_{23}$ rotational
transition of ortho-H$_2$O.  This transition collisionally inverts
over a wide range of conditions: temperatures of several hundred K,
abundances $n_{\rm{H_2O}}/n_{\rm{H_2}} \sim 2$--$4\times10^{-4}$
\citep{Coo85} and densities up to $n_{\rm{H_2}}
\sim 10^{11}$ cm$^{-3}$ where collisions thermalise rotational energy
levels \citep{Coh89}.  Model calculations \citep{Coo85} predict that
H$_2$O masers are found between two well-defined radii in the
circumstellar envelopes. The outer radius is determined by collision
rates that are too low to invert the masing energy levels, and the
inner radius by high densities that destroy the inversion through too
many collisions.  The masing region is predicted to occur at larger
radii for stars with higher mass-loss rates, from $10^{14}$ cm for
$\dot{\rm{M}} = 10^{-7}$ M$_{\odot}$ yr$^{-1}$ to $\geq 10^{15}$ cm
for $\dot{\rm{M}} = 10^{-5}$ M$_{\odot}$ yr$^{-1}$.  This model has
been confirmed observationally for Miras
\citep{Lan87,Bai03b}.  The dissociation of H$_2$O by external UV
radiation into OH and H is important at radii beyond $10^{15}$ cm and
H$_2$O maser emission is absent in stars with mass-loss rates below
$10^{-8}$ M$_{\odot}$ yr$^{-1}$ \citep{Hab96}.

Very short-period Miras (P $<$ 200 days) and semi-regular variables
rarely have H$_2$O masers, due to their low mass-loss rates
\citep{Ben96,Szy97}.  For those Miras and semi-regulars with H$_2$O
masers, the spectra are generally narrow and at velocities close to
that of the central star, often in regions where circumstellar
material is strongly accelerated outwards and tangential amplification
dominates over radial amplification \citep{Men91,Eng96,Eng97a}.
Interferometric observations of Miras and semi-regulars \citep[][and
references therein]{Cha95,Bai03b} confirm that the H$_2$O masers are
generally located in a thick shell in a radially accelerating region
where the envelope expansion velocity increases with increasing
distance from the central star. In contrast, OH/IR stars have wider
H$_2$O spectra, often double-peaked and comparable to the OH maser
velocity widths \citep{Tak94,Eng96,Szy97}.  Double-peaked spectra
occur from H$_2$O masers located in regions of the circumstellar
envelope where the velocity gradients are small and radial
amplification dominates.  The velocity separation is also larger in
stars with higher mass-loss rates as these stars have larger H$_2$O
maser regions.  The blue-shifted peaks are usually stronger than the
red-shifted peaks \citep{Tak94,Eng96}, probably due to absorption of
red-shifted photons as they pass through the high-density region
around the star inside the H$_2$O maser zone.

In the few previous studies of stellar H$_2$O masers which included
post-AGB stars, a rapid decrease in detection rate was found 
as stars leave the AGB \citep{Eng96,Tak01,Val01}.  Approximately
10--20 post-AGB sources with H$_2$O maser emission have been
detected.  In the first study there was evidence for an evolution from
double-peaked profiles back to Mira-like single-peaked profiles in
older post-AGB stars which still retained H$_2$O masers.  This was
considered to be due to decreasing mass-loss rates after the AGB and a
consequent switch back to tangential pumping of the H$_2$O masers.
The trend of the blue-shifted features in double-peaked profiles being
brighter than the red-shifted features, as in OH/IR stars, is
continued in post-AGB stars.

H$_2$O maser emission has been detected from only two PN; K 3--35
\citep[IRAS 19255$+$2123,][]{Mir01} and IRAS 17347$-$3139
\citep{Deg04b}. In K 3--35 the H$_2$O masers are located near the
central star, probably in the inner part of a rotating and expanding
torus, and far from the star at the tips of $\sim 800$ year old jets
that are traced by radio continuum and optical spectral-line emission.
In IRAS 17347$-$3139 they are likely to be in a similar torus close to
the central star.

H$_2$O maser emission has also been detected from five post-AGB
stars. In all cases the spectra reveal high-velocity outflows. For
four of the five sources high resolution imaging observations have
shown high-velocity, collimated jets traced by the H$_2$O masers. The
nature and properties of these `water-fountain' sources are discussed
in Section \ref{Sec:high_velocity_water}.

\subsection{SiO Maser Emission in Evolved Stars} \label{Sec:sio_late_type}

SiO masers are primarily observed in evolved stars including Miras and
OH/IR stars, where they arise from collisionally and/or
radiatively pumped rotational levels of excited vibrational states.
The transitions most studied are the $v = 1$, $J = 1$--0 and $v = 2$,
$J = 1$--0 transitions near 43 GHz, and the $v = 1$, $J = 2$--1
transition near 86 GHz.  The SiO masers require column densities of
$10^{18}$--$10^{19}$ cm$^{-2}$ and temperatures above $\sim 1200$ K
\citep{Eli92a}.  There is little correlation between the presence of 
SiO masers and mass-loss rate or envelope expansion velocity,
indicating that these masers form inside the dust condensation radius.
Outside the dust condensation radius, most SiO is locked in dust
grains and is no longer available for masing transitions.

The SiO maser emission from Mira variables and OH/IR stars is
generally seen in ring-like structures \citep{Hab96}, indicating that
they are tangentially amplified in a spherical or quasi-spherical
shell.  The spectra commonly show a broad peak that is 5--15 km
s$^{-1}$ wide and centred on the stellar velocity \citep{Nym98}.  The
emission is highly variable, with lifetimes for individual clumps of
no more than a few months seen in both spectral and high-resolution
observations \citep{Eli92a,Hum02}.  Both infalling and outflow motions
have been seen, most notably in multiple observations of the 43 GHz
SiO masers around the Mira variable TX Cam made using the Very Long
Baseline Array (VLBA) \citep{Dia03}.  The time variability and chaotic
velocity field reflect the conditions in the upper atmospheres of these
stars.

Detection rates for SiO masers are very low for post-AGB stars. From a
large study of SiO maser emission in evolved stars, \citet{Nym98}
detected only one post-AGB source (at 43 GHz), the bipolar PPN OH
231.8$+$4.2.  This has an unusual double-peaked spectrum with two
features separated by 12~km~s$^{-1}$.  This source is a candidate
binary, with the H$_2$O and 43~GHz SiO masers associated with a known
Mira variable \citep{Gom01,San02} and a hotter companion star that may
be driving a fast wind that shapes the mass loss from the Mira.  The
SiO masers may be in a rotating and infalling torus with a radius of 6
AU located around the Mira.  IRAS 19312$+$1950 is another post-AGB SiO
maser source with a double-peaked profile at 43 GHz \citep{Nak00}.
The peaks are separated by 26 km s$^{-1}$ and a possible
interpretation is that the emission comes from a rotating or expanding
circumstellar shell.

\section{SAMPLE SELECTION} 
 
For the OH observations, a complete sample of 88 sources was selected
from the ATCA/VLA survey as likely post-AGB sources on the basis of
their IRAS and MSX far-infrared colours.  Three of the sources (d29,
b262, v280) from the initial selection were found to be star-forming regions
\citep[][and Paper I]{Zij90} and were not observed at 22 GHz, giving a
sample size of 85 in this paper.

\begin{figure} 
\centering
\rotatebox{270}{\scalebox{0.5}{\plotone{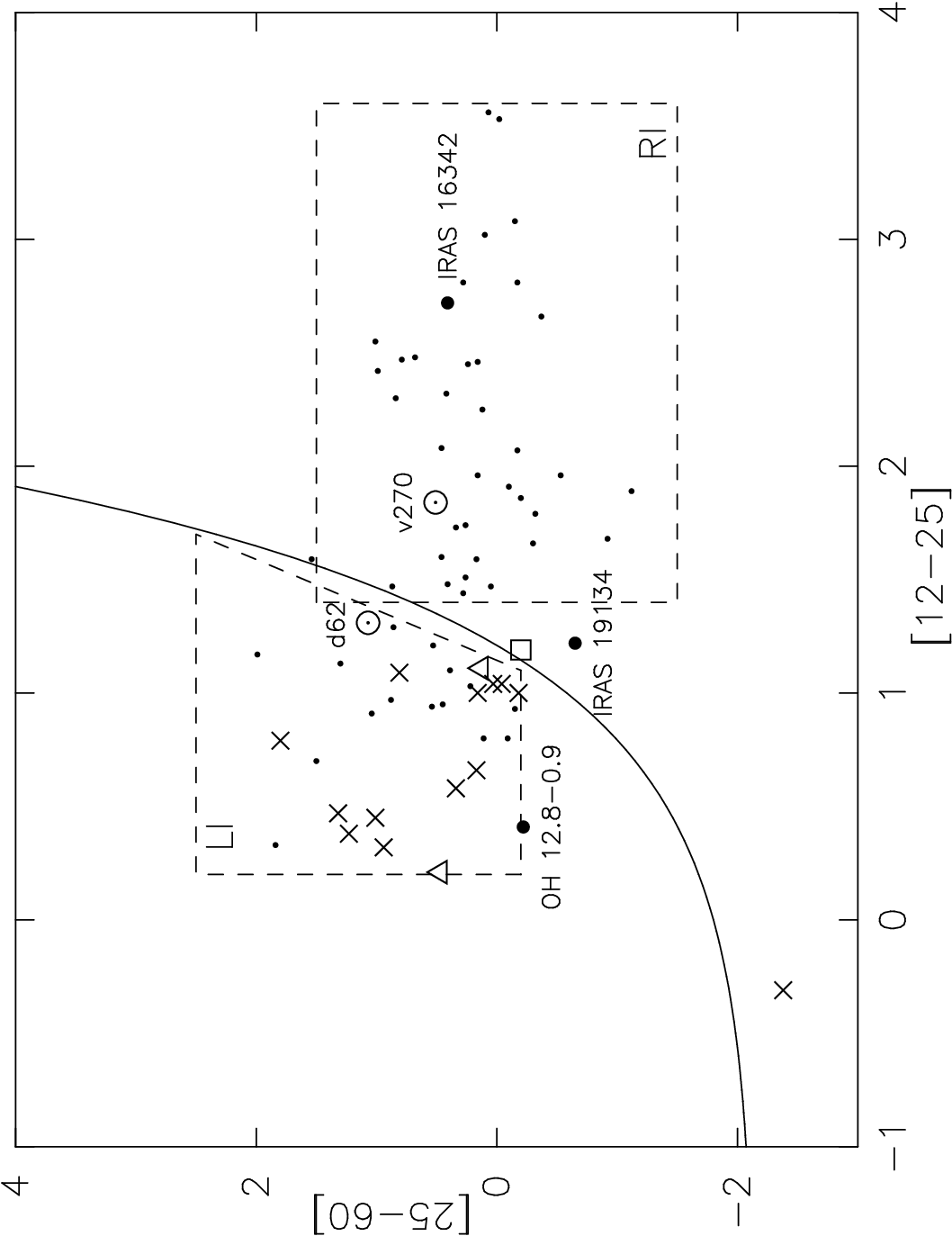}}}
\rotatebox{270}{\scalebox{0.5}{\plotone{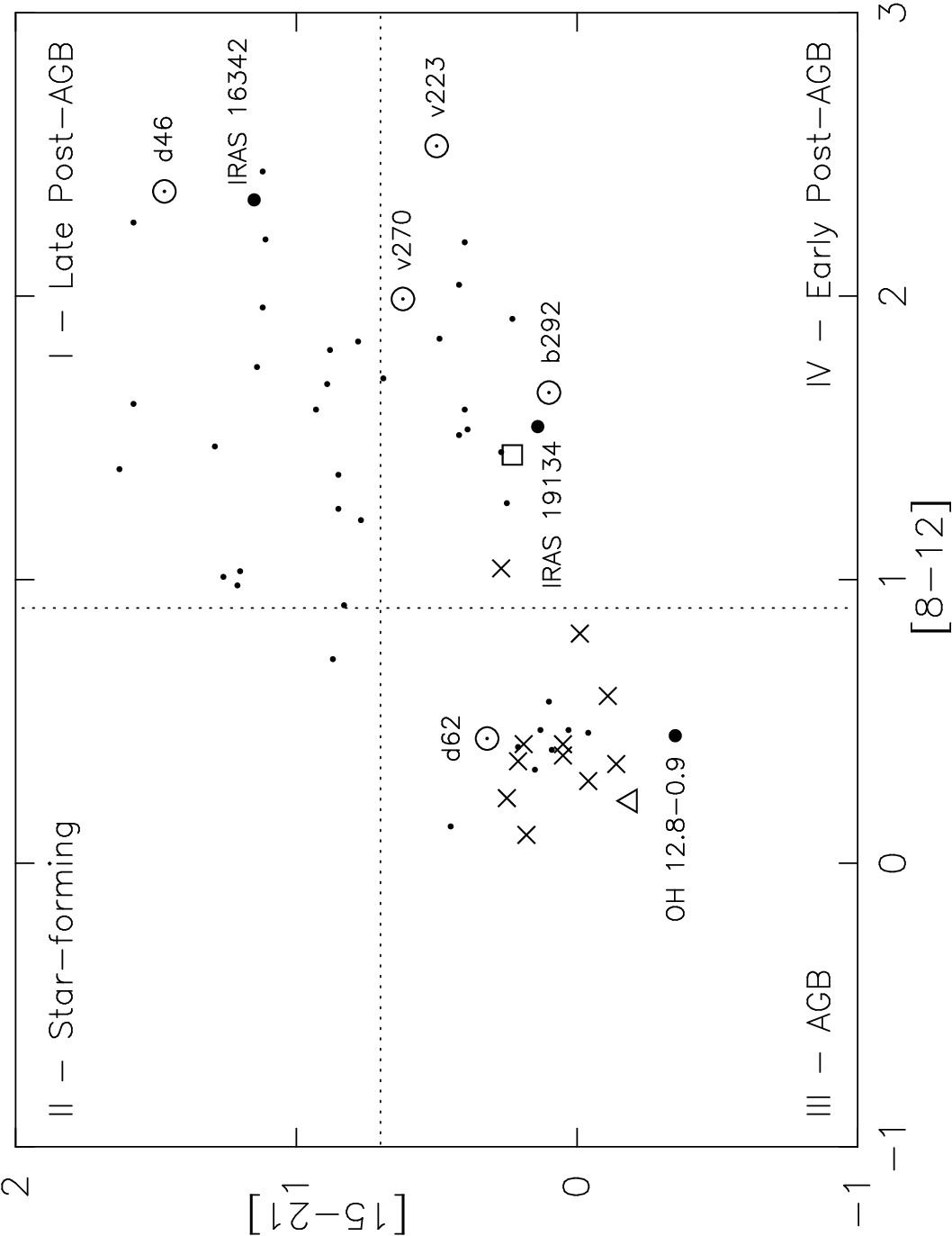}}}
\figcaption{IRAS ({\it top}) and MSX ({\it bottom}) colour-colour plots
of the 85 sources in the post-AGB sample.  Sources with no detected
H$_2$O maser emission are indicated with {\it dots}.  Sources with
regular spectra (`R') are indicated with {\it crosses}.  Those with
irregular (`I') spectra are marked with {\it triangles}.  The one
source with a narrow profile at the stellar velocity (`S') is
indicated with a {\it square} and {\it dotted circles} mark sources
with high-velocity features (`H'). Sources with high-velocity H$_2$O
maser emission are labelled. The three high-velocity sources, IRAS
16342$-$3814, IRAS 19134$+$2131 and OH12.8$-$0.9, that have previously
published detections, are shown as {\it filled circles} (Section
\ref{Sec:high_velocity_water}) which are labelled.
\label{Fi:IR_water}}
\end{figure}

Figure \ref{Fi:IR_water} plots the sources against their IRAS
[12--25], [25--60] and MSX [8--12], [15-21] colours, where [a--b] $=
2.5$ log (S$_b$/S$_a$), S is the flux density in Jy and a, b, are
wavelengths in $\mu$m.  The sources detected at 22 GHz are plotted
with various symbols indicating the different profile types, as
described in the caption.  The 22 GHz results are discussed in
Section \ref{Sec:water_results}.

In the IRAS plot the evolutionary path for AGB stars defined by \citet{Van88} 
is shown.  The regions associated with more and less
massive post-AGB stars respectively are denoted `LI' (Left of IRAS)
and `RI' (Right of IRAS) \citep{Sev02a}.  The LI sources are
assoicated with the most massive AGB stars, with an average mass of 
4 $M_{\odot}$, as determined from their Galactic scale heights
\citep{Sev02b}. These stars have very high mass-loss rates and
correspondingly high 60 $\mu$m excesses. \citet{Van97} found that the
LI stars may have an earlier AGB termination than less massive stars,
and the stars in the LI region may be at a turn-around point in their
evolution before they evolve to redder IRAS [12--25] colours
\citep{Sev02b}.  From Paper I, LI stars typically have regular OH maser
profiles. At 1612 MHz, only 3 out of 30 LI
sources have non-regular profiles.  They also have a much lower
detection rate for OH mainline masers with OH 1667 MHz emission detected from 
just 14 sources and only one detection at 1665 MHz (Paper I).

The LI region overlaps with the Quad III (AGB) region in the MSX
diagram (next paragraph).  Coupled with their regular OH
profiles and high mass-loss rates, this suggests they may still be AGB
stars.  The nature of LI sources will be discussed more in Section
\ref{Sec:LI_are_AGB}.

The RI region encompasses sources with IRAS colours that have
traditionally been associated with post-AGB stars, PPN and PN
\citep{Van89}.  In comparison with LI sources, the RI sources have an
average mass of 1.7 M$_{\odot}$ and lower OH maser outflow velocities
\citep{Sev02b}, and more OH mainline emission (Paper I).  RI sources 
also have more irregular OH maser profiles than LI sources (Paper I),
indicating they may be more evolved.

The MSX plot is divided into four quadrants.  Increasing MSX [8--12]
colour was found to accurately indicate evolution of stars off the
AGB, and increasing MSX [15--21] colour indicates further post-AGB
evolution and possibly the onset of the fast wind \citep{Sev02a}.  Thus,
older post-AGB stars are found in Quad I, and young post-AGB stars in
Quad IV \citep{Sev02a}.  The other two quadrants, II and III, contain
star-forming regions and AGB stars respectively.  The MSX Quad I and
IV sources overlap with the IRAS RI region. However, the MSX colours
provide better selection criteria for separating younger and older
post-AGB sources.

Table \ref{Ta:water_source_list} lists the 85 sources included in the
sample.  The first four columns give the source identifier from
the ATCA/VLA survey, the IRAS name and the source classifications as
RI, LI, Quad I or Quad IV (note some sources have two
classifications).  The final two columns indicate whether 22 GHz
H$_2$O maser emission and 86 GHz SiO maser emission were detected
(y/n).  A blank entry in the SiO column indicates the source was
not observed.  See \citet{Sev02a,Sev02b} and Paper I
for further details on the selection criteria.

\section{OBSERVATIONS} 
 
\subsection{H$_{\mathbf{2}}$O Maser Observations}

22 GHz observations for all 85 sources were taken as service
observations with the Tidbinbilla 70-m radio telescope between 2003
April and 2004 May.  The rest frequency assumed is 22.23507985~GHz,
which is the weighted average of the three strongest hyperfine
components of this transition \citep{Mor73}.  No other hyperfine
splitting effects are taken into account.  The beam size at 22 GHz is
0.8~arcmin.  The bandpass was 16~MHz with 8193 channels, and one
circular polarisation (left-hand circular polarisation) was
recorded.  A four-point position-switching mode was used, with
observations east-on-on-west relative to the source position.  Offsets
were 10~arcmin with 3.5~mins spent at each position, giving 7 mins
on-source total. Calibration observations to correct the pointing were
performed as necessary, giving an average positional error better than 7
arcsec.  Most sources were observed once, excepting d46 and d47, which
were observed two and three times respectively.  The weak sources d62,
d103 and v149 were also observed twice to confirm the detection and
the spectra were averaged as they did not change significantly.  Table
\ref{Ta:water_obs_sched} lists the UT dates of observations and the
range of system temperatures ($T_{\rm{sys}}$) for each date.  The
system temperatures are elevation and weather dependent.
 
Initial data reduction to calculate quotient spectra and to average
the two on-source observations was carried out using SPC, a
single-dish spectral line reduction package supported by the Australia
Telescope National Facility (ATNF).  Further reduction including
velocity, baseline and flux calibration, smoothing and flagging was
carried out with the Spectral Line Analysis Package
\citep[SLAP;][]{Sta85}. All velocities are given relative to the Local
Standard of Rest (radio definition).
 
The 16~MHz bandpass gives a velocity coverage of approximately 200 km
s$^{-1}$ and a channel separation of 0.026~km~s$^{-1}$.  All spectra
were smoothed with a Gaussian function in SLAP to 0.18~km~s$^{-1}$
velocity resolution to enable an accurate comparison of spectral
features and peak fluxes with the OH maser observations from Paper I.
This resulted in rms errors of between 0.04 and 0.13~Jy~beam$^{-1}$ in
the final spectra.

\subsection{Mopra 86~GHz SiO Observations}
 
A subset of 11 sources (d46, d47, d56, d168, d189, b11, b34, b292,
b301, v67, and v117) were observed using the Mopra radio telescope
near Coonabarabran, NSW, which is operated jointly by ATNF and the
University of New South Wales as a National Facility.  Observations
were taken in 2004 June and August using a 3-mm SIS receiver system
and a 2-bit Australia Telescope correlator as the backend. Spectra
were taken at SiO maser line rest frequency 86.243442~GHz. The sources
are included in Table \ref{Ta:water_obs_sched}.  Two linear
polarisations were used with 1024 channels over a bandwidth of 64~MHz,
giving a velocity resolution of 0.22~km~s$^{-1}$.  Doppler tracking
was implemented. Unfortunately, technical issues and adverse weather,
including snow, severely limited the time available for the SiO
observations, resulting in a restricted number of sources being
measured.

The observations were performed in an `on-off' manner, with the
telescope on-source for 1 min followed by a sky observation of the
same length.  This cycle was repeated 10 times over a 20~min
observation.  Paddle measurements (where a room-temperature load is
driven in front of the receiver) were performed about once an hour to
measure the system temperature between observations. However, due to
rapidly changing atmospheric conditions and system temperatures, it
was not possible to perform reliable amplitude calibrations.

Nearby strong SiO masers were observed once per hour to calibrate the
pointing to within 6~arcsec. For the on-source observing time of
10~mins, the system sensitivity (1$\sigma$ rms) was 90 to 220~mK per
channel for $T_{\rm{sys}}$ measurements ranging from 200 to 500~K.
Most sources were observed two or three times and the observations
averaged together.

Initial data reduction to calculate quotient spectra, average spectra,
and set the velocity scale relative to the Local Standard of Rest
was done using SPC. Spectra were then converted to
the SLAP format and final plots were made with SLAP. Due to a problem
with data headers, the velocities of the SiO spectra are uncertain to
within $\pm$ 1 km s$^{-1}$.

\subsection{Radio Continuum Observations of d46}

Radio continuum observations of d46 were taken in November 1998 with
the Australia Telescope Compact Array (ATCA) at 3, 6 and 13 cm. The central
observing frequencies were 8640, 4800 and 2496 MHz respectively.  The
observations were taken using the 6D array with a bandwidth of 128
MHz, and a longest baseline of 6 km. At 3 and 6 cm short-track
observations were taken with a total observing time of 30 minutes,
taken as four cuts of 7--8 minutes. At 13 cm the on-source observing
time was 55 mins, taken as 11 cuts over 12 hours. Observations of a
secondary calibrator source was taken to correct for atmospheric
amplitude and phase variations. The absolute flux scale was determine
from observations of PKS B1934$-$638. The data were reduced using the radio
astronomy package Miriad \citep{Sau95}.

\begin{deluxetable}{l l l c c c}
\tabletypesize{\small}
\tablewidth{0pt}
\tablecaption{Source list with H$_2$O and SiO maser detections. \label{Ta:water_source_list}} 
\tablecolumns{6}
\tablehead{ 
ID & IRAS Name & \multicolumn{2}{c}{Classification} & 
H$_2$O  & SiO \\
& & \multicolumn{1}{l}{{\fns MSX}} & \multicolumn{1}{l}{{\fns IRAS}}}
\startdata
d3   &  14341$-$6211 & Q I  & RI   & n & \nodata \\ 
d34  &  15338$-$5202 & Q I  &      & n & \nodata \\ 
d39  &  15367$-$5420 &      & LI   & n & \nodata \\ 
d46  &  15445$-$5449 & Q I  &      & y & n \\ 
d47  &  15452$-$5459 &      & LI   & y & y \\ 
d56  &  15514$-$5323 &      & LI   & n & n \\ 
d62  &  15544$-$5332 &      & LI   & y & \nodata \\ 
d93  &  16209$-$4714 &      & RI   & n & \nodata \\ 
d103 &  16314$-$5018 & Q IV &      & y & \nodata \\ 
d117 &  16372$-$4808 & Q I  &      & n & \nodata \\ 
d150 &  16507$-$4810 &      & RI          & n & \nodata \\ 
d168 &  17004$-$4119 &      & LI          & y & y \\ 
d189 &  17088$-$4221 & Q IV &     & y & n \\ 
b5   &  17097$-$3624 &      & RI          & n & \nodata \\ 
d190 &  17103$-$3702 &      & RI          & n & \nodata \\ 
b11  &  17150$-$3224 & Q I  &  RI  & n & n \\ 
d197 &  17151$-$3845 & Q I  &      & n & \nodata \\ 
b14  &  17164$-$3226 & Q I  &  RI  & n & \nodata \\ 
b15  &  17162$-$3751 & Q I  &      & n & \nodata \\ 
b17  &  17168$-$3736 & Q IV &  RI & n & \nodata \\ 
d200 &  17188$-$3838 &      & LI          & y & \nodata \\ 
b25  &  17193$-$3546 & & LI          & y & \nodata \\ 
b30  &  17205$-$3556 & & LI          & n & \nodata \\ 
b31  &  17207$-$3632 & & LI          & n & \nodata \\ 
b33  &  17227$-$3623 & Q IV &     & n & \nodata \\ 
b34  &  17230$-$3348 & Q IV &  LI & n & n \\ 
d202 &  17245$-$3951 & Q I &  RI  & n & \nodata \\ 
b44  &  17256$-$3258 & & LI          & n & \nodata \\ 
b62  &  17293$-$3302 &  & RI         & n & \nodata \\ 
b68  &  17310$-$3432 & & RI          & n & \nodata \\ 
b70  &  17317$-$2743 & & RI          & n & \nodata \\ 
b96  &  17359$-$2902 & Q I &  RI  & n & \nodata \\ 
b106 &  17367$-$3134 & Q I &      & n & \nodata \\ 
b112 &  17370$-$3357 & & RI          & n & \nodata \\ 
b114 &  17371$-$2747 & & RI          & n & \nodata \\ 
b128 &  17385$-$3332 & & RI          & n & \nodata \\ 
b130 &  17390$-$2809 & Q IV &     & n & \nodata \\ 
b133 &  17392$-$3020 & & LI          & y & \nodata \\ 
b134 &  17393$-$2727 & & RI          & n & \nodata \\ 
b143 &  17404$-$2713 & & RI          & n & \nodata \\ 
b155 &  17414$-$3108 & &  LI         & n & \nodata \\ 
b165 &  17426$-$2804 & & LI          & n & \nodata \\ 
b199 &  17461$-$2741 & Q IV &     & n & \nodata \\ 
b209 &  17479$-$3032 & & RI          & n & \nodata \\ 
b210 &  17482$-$2501 & & RI          & n & \nodata \\ 
b228 &  17506$-$2955 & & RI          & n & \nodata \\ 
b246 &  17543$-$3102 & & RI          & n & \nodata \\ 
b250 &  17548$-$2753 & & RI          & n & \nodata \\ 
b251 &  17550$-$2120 & Q IV &  RI & n & \nodata \\ 
b258 &  17560$-$2027 & & RI          & n & \nodata \\ 
b263 &  17576$-$2653 & Q I &  RI  & n & \nodata \\ 
b266 &  17582$-$2619 & Q I &  RI  & n & \nodata \\ 
b292 &  18043$-$2116 & Q IV &     & y & n \\ 
b300 &  18051$-$2415 & Q IV &  RI & n & \nodata \\ 
b301 &  18052$-$2016 & & LI          & y & y \\ 
b304 &  18070$-$2332 & Q IV &     & n & \nodata \\ 
v41  &  18076$-$1853 & Q I  &     & n & \nodata \\ 
v45  &  18087$-$1440 & Q I &  RI  & n & \nodata \\ 
v50  &  18092$-$2347 & & LI          & n & \nodata \\ 
v53  &  18100$-$1915 & & LI          & n & \nodata \\ 
v56  &  18103$-$1738 & & LI          & y & \nodata \\ 
v67  &  18135$-$1456 & Q IV &  RI & n & n \\ 
v87  &  18182$-$1504 & & LI          & n & \nodata \\ 
v117 &  18246$-$1032 & & RI          & n & n \\ 
v120 &  18257$-$1052 & & LI          & n & \nodata \\ 
v121 &  18257$-$1000 & & LI          & y & \nodata \\ 
v132 &  18276$-$1431 & Q I &  RI  & n & \nodata \\ 
v146 &  18310$-$0806 & Q I&       & n & \nodata \\ 
v149 &  18314$-$0900 &      & LI            & y & \nodata \\ 
v154 &  18327$-$0715 & & LI          & y & \nodata \\ 
v162 &  18342$-$0655 & Q I &      & n & \nodata \\ 
v169 &  18355$-$0712 & & RI          & n & \nodata \\ 
v172 &  18361$-$0647 & & LI          & n & \nodata \\ 
v189 &  18389$-$0601 & & LI          & n & \nodata \\ 
v204 &  18420$-$0512 & Q I &  RI  & n & \nodata \\ 
v211 &  18432$-$0149 & & LI          & n & \nodata \\ 
v212 &  18434$-$0202 & & LI          & y & \nodata \\ 
v223 &  18450$-$0148 & Q IV &     & y & \nodata \\ 
v228 &  18460$-$0254 & & LI          & y & \nodata \\ 
v231 &  18467$-$0238 & Q IV &  RI & n & \nodata \\ 
v237 &  18485$+$0642 & Q I &  RI  & n & \nodata \\ 
v239 &  18488$-$0107 & & LI          & y & \nodata \\ 
v268 &  18588$+$0428 & & LI          & y & \nodata \\ 
v270 &  18596$+$0315 & Q IV &  RI & y & \nodata \\ 
v274 &  19024$+$0044 & Q I &  RI  & n & \nodata \\ 
\enddata
\end{deluxetable}

\begin{deluxetable}{l c}
\tablewidth{0pt}
\tablecolumns{2}
\centering
\tablecaption{Schedule of H$_2$O maser observations \label{Ta:water_obs_sched}} 
\tablehead{\colhead{UT date} & \colhead{$T_{\rm{sys}}$ (K)}} 
\startdata
\cutinhead{Tidbinbilla, 22 GHz}
2003 Apr 23 & 77 \\ 
2003 Jul 1& 55--90 \\ 
2003 Aug 28 & 56 \\ 
2003 Nov 15 & 82--100 \\ 
2004 Feb 27--28 & 59--122 \\ 
2004 Mar 4 & 45--82 \\ 
2004 Mar 9 & 51--76 \\ 
2004 Mar 10 & 61--81 \\ 
2004 Mar 11--12 & 91--182 \\ 
2004 Mar 16--17 & 63--119 \\ 
2004 Apr 14 & 76 \\ 
2004 May 16 & 35--56 \\ 
2004 May 20 & 50--60 \\
\cutinhead{Mopra, 86 GHz} 
2004 June 18 & 200--500 \\
2004 August 26 & 200--500 
\enddata
\end{deluxetable}

\section{RESULTS} \label{Sec:water_results}
 
From the Tidbinbilla observations, 21 stars from the sample of 85
were detected at 22~GHz. Summary data are shown in Table
\ref{Ta:water_results}, with the columns indicating:
 
\begin{itemize} 
\item{1: Source name as in the ATCA/VLA survey.  Repeat observations  
are designated by bracketed numbers.  Sources for which two
observations have been averaged are indicated with a superscript `a'.}

\item{2-4: Velocity, peak flux density and integrated flux density for 
the blue-shifted emission for sources with clearly separated blue- and
red-shifted peaks, or for the strongest emission feature for sources
with one-sided or irregular spectra.}

\item{5-7: Velocity, peak flux density and integrated flux density for 
the red-shifted feature of double-peaked sources.}

\item{8-9: The most extreme blue and red velocities for which emission was 
detected.}

\item{10: H$_2$O maser profile type.  `R' indicates a regular 
double-peaked spectrum, or a single-peaked spectrum where the single
feature is near the most red- or blue-shifted OH peaks or velocity
limit. `S' indicates a spectrum with a narrow profile centred near the
stellar velocity.  `I' indicates an irregular spectrum with many
emission features over a velocity range similar to the OH velocity
range.  `H' indicates a spectrum with features at significantly larger
velocities than the OH emission, relative to the central stellar
velocity.}

\end{itemize} 

\begin{figure} 
\centering
\scalebox{0.819}{\plotone{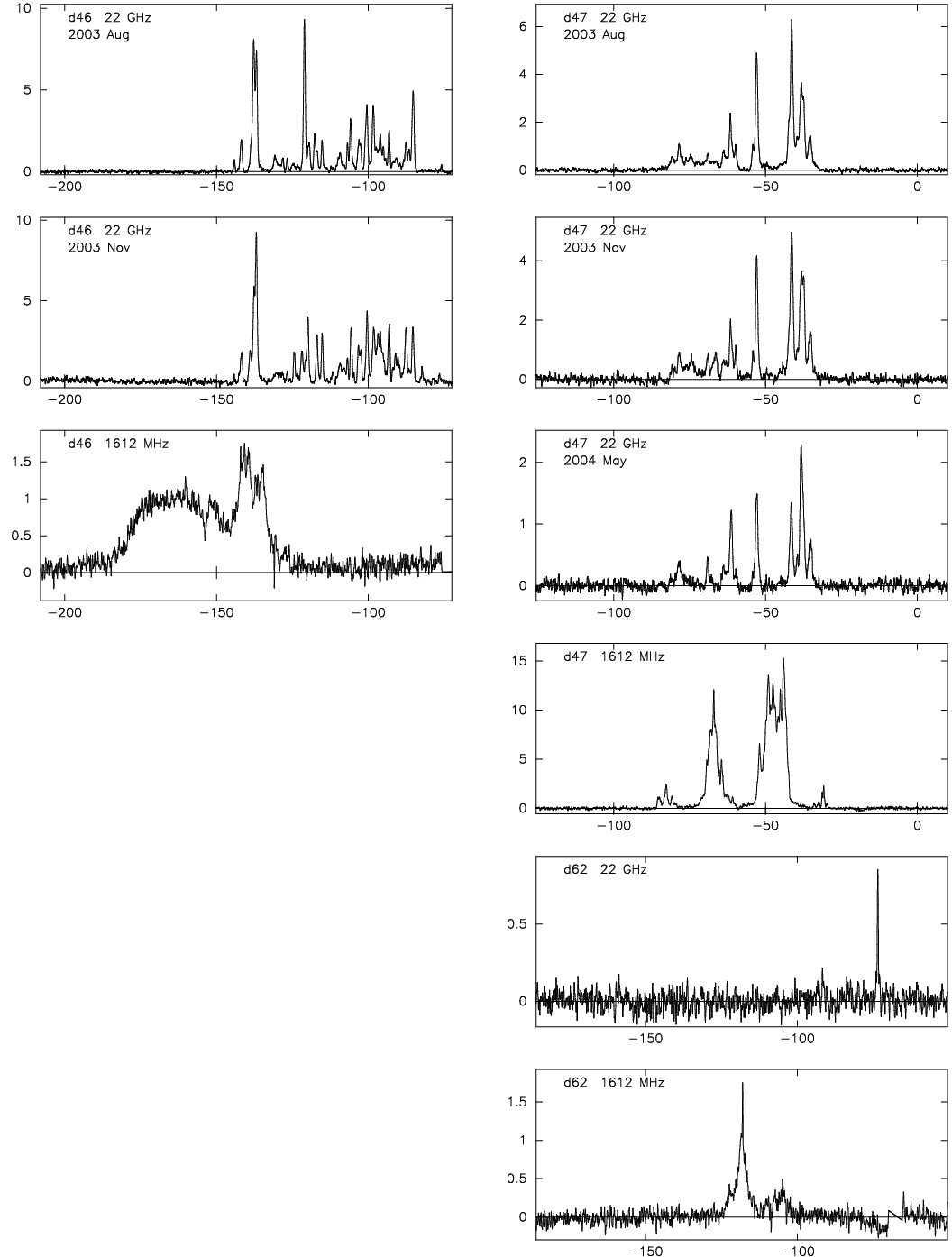}}
\figcaption{22~GHz H$_2$O maser spectra from Tidbinbilla telescope
observations. Source names are noted on individual spectra. For d46
and d47 the observing dates are also given.  The OH 1612 MHz maser
spectra from Paper I are plotted for all the sources, for comparison.
\label{Fi:H2O_thumbnails:a}}
\end{figure}

\begin{figure} 
\centering 
\scalebox{0.819}{\plotone{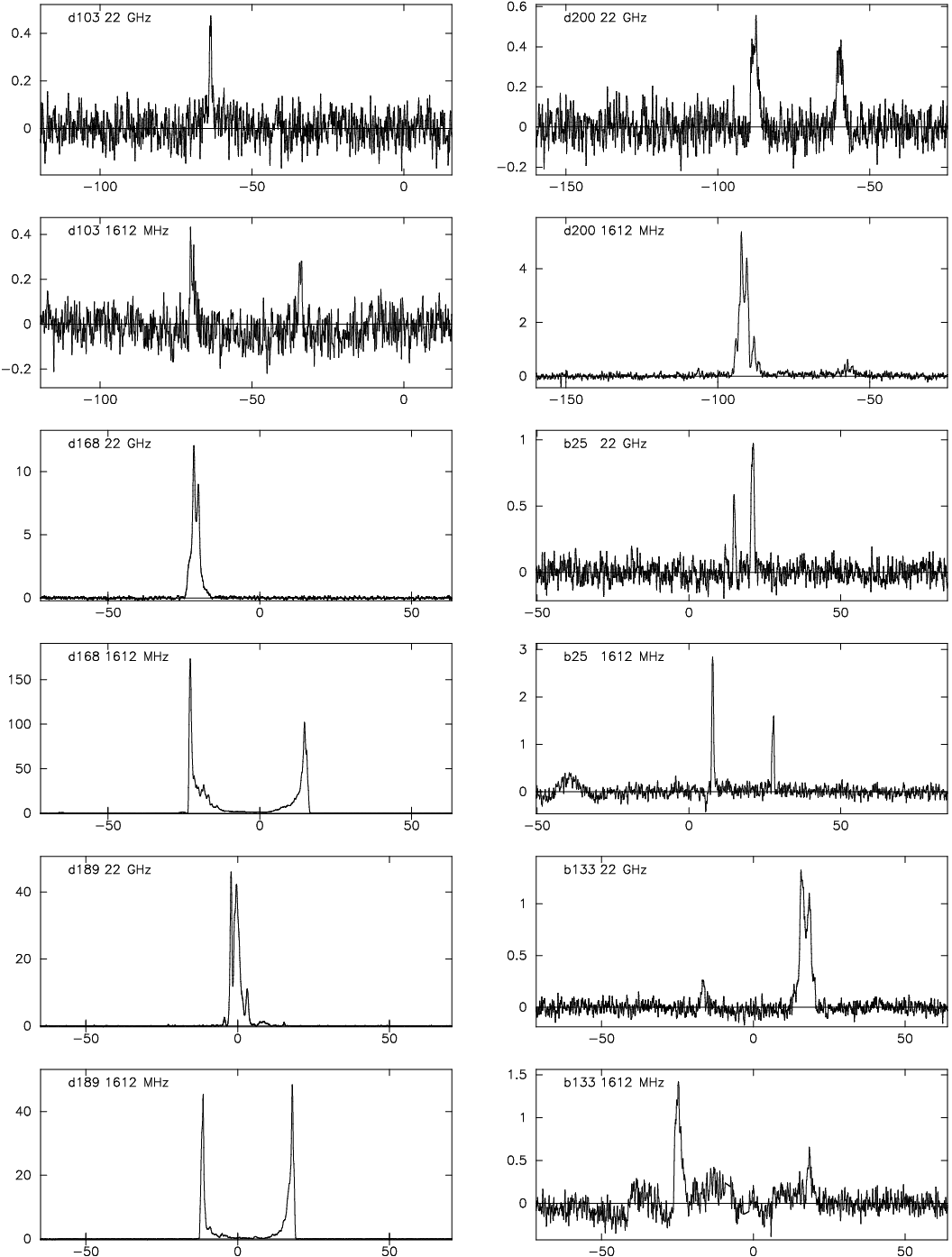}}
\figcaption{22~GHz spectra continued. \label{Fi:H2O_thumbnails:b}}
\end{figure}

\begin{figure} 
\centering 
\scalebox{0.819}{\plotone{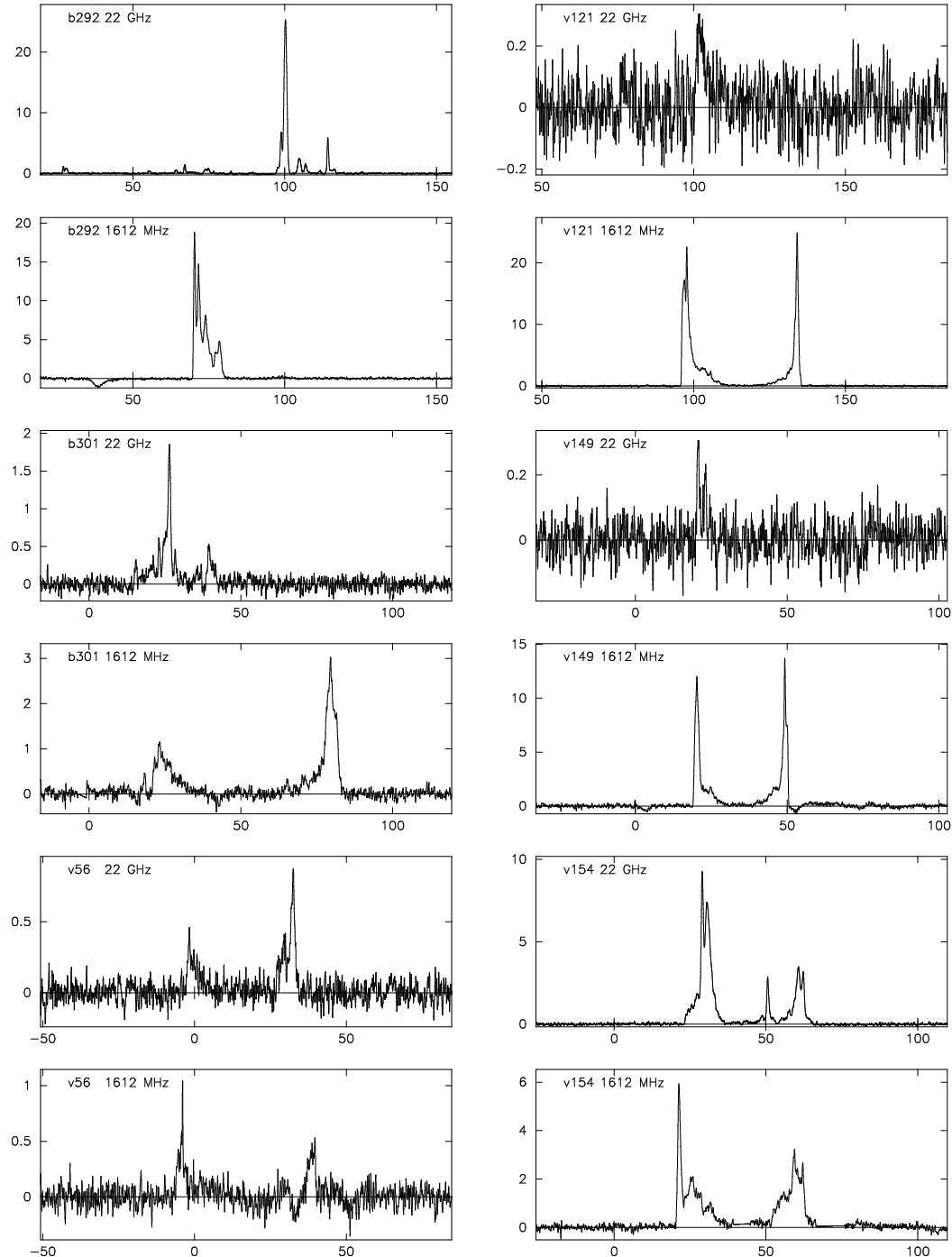}} 
\figcaption{22~GHz spectra continued.  Note that for b292 the 
velocity range plotted does not cover the full extent of the maser
emission (see Figure \ref{Fi:b292_v223}). \label{Fi:H2O_thumbnails:c}}
\end{figure}

\begin{figure} 
\centering 
\scalebox{0.819}{\plotone{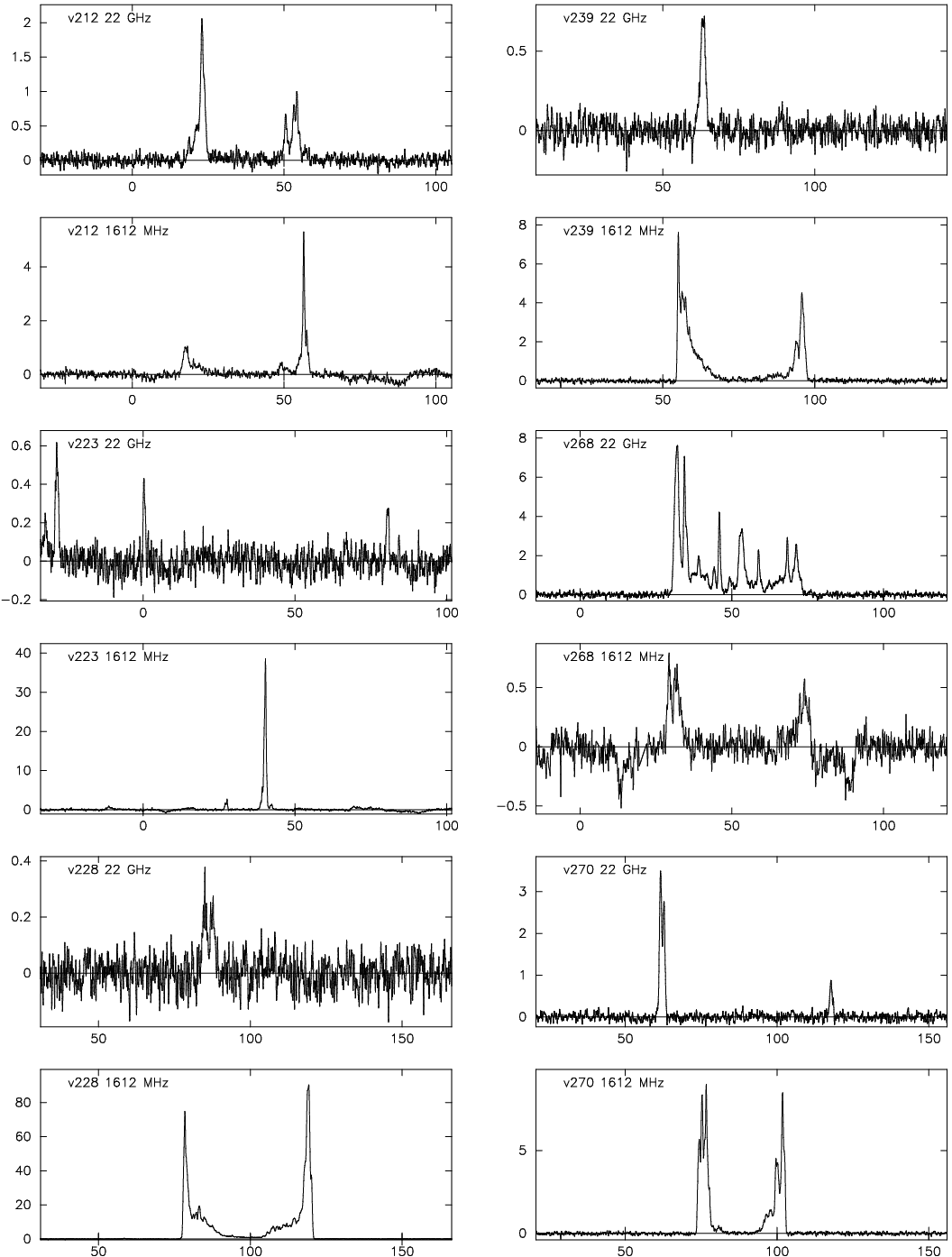}} 
\figcaption{22~GHz spectra continued.  Note that for v223 the 
velocity range plotted does not cover the full extent of the maser
emission (see Figure \ref{Fi:b292_v223}). \label{Fi:H2O_thumbnails:d}}
\end{figure}

\begin{figure}
\centering
\rotatebox{270.0}{\scalebox{0.65}{\plotone{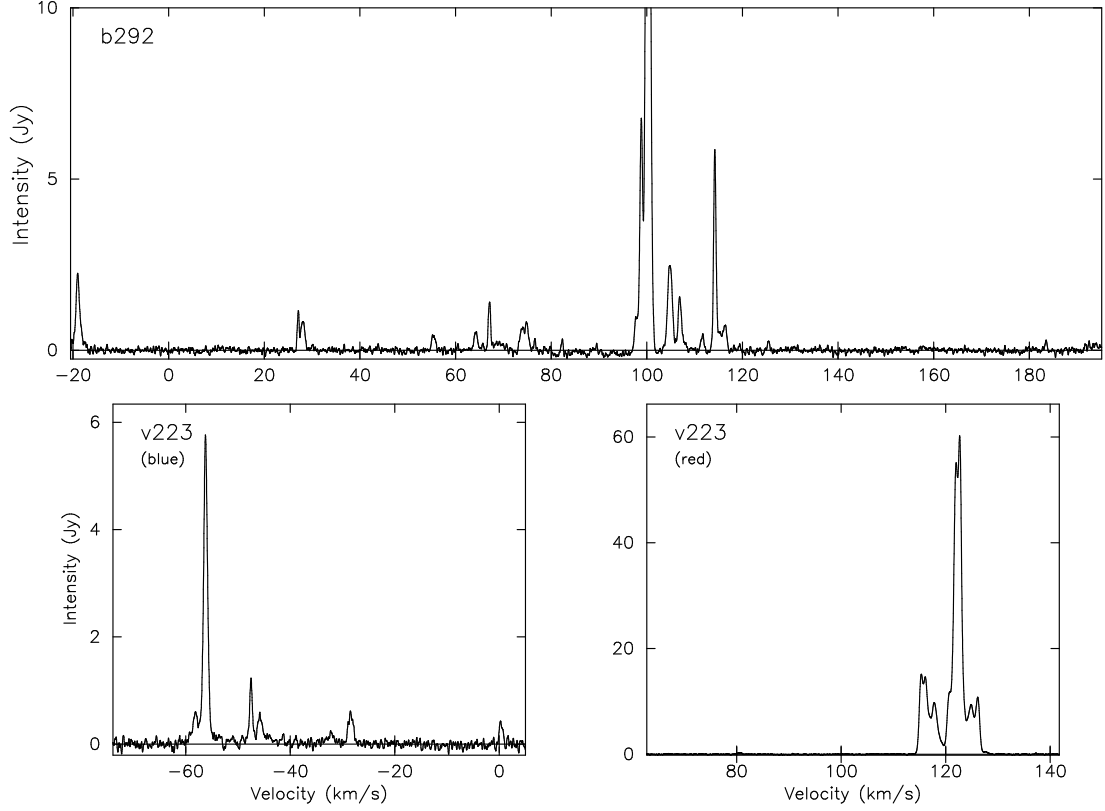}}}
\figcaption{H$_2$O maser spectra for b292 and v223 showing the high
velocity features.  The {\it top} subfigure showing the spectrum for
b292 over the entire observed velocity range.  The stellar velocity is
87.3~km~s$^{-1}$.  The vertical axis is truncated to enable the weaker
high-velocity features to be seen more easily. The full peak at
100~km~s$^{-1}$ can be seen in Figure \ref{Fi:H2O_thumbnails:c}. {\it
Bottom:} The most blue-shifted ({\it left}) and red-shifted ({\it
right}) features in the v223 spectra, on different vertical scales.
The central velocity range of the v223 spectrum is shown in Figure
\ref{Fi:H2O_thumbnails:d}.  The stellar velocity is 33.9~km~s$^{-1}$.
\label{Fi:b292_v223}}
\end{figure}

Figures \ref{Fi:H2O_thumbnails:a} to \ref{Fi:H2O_thumbnails:d} show
the Gaussian smoothed spectra for the detections at 22~GHz, including
the repeated observations of d46 and d47. The OH 1612 MHz spectra from
Paper I are also shown for comparison.  Figure \ref{Fi:b292_v223}
shows the H$_2$O maser spectra of b292 and v223 for the full velocity
range observed, revealing the very high-velocity features present in
these two sources.

Table \ref{Ta:water_stats} gives the number of OH 1612 MHz detections
and 22 GHz H$_2$O detections for each of the four infrared selection
groups and for the four spectral profile types R, S, I and H.  The
columns give, from left to right, the frequency or profile type
referred to, the total number of sources detected at that frequency or
profile, and the number of RI, LI, Quad~I and Quad~IV sources
detected.

\begin{figure}
\centering
\scalebox{0.45}{\plotone{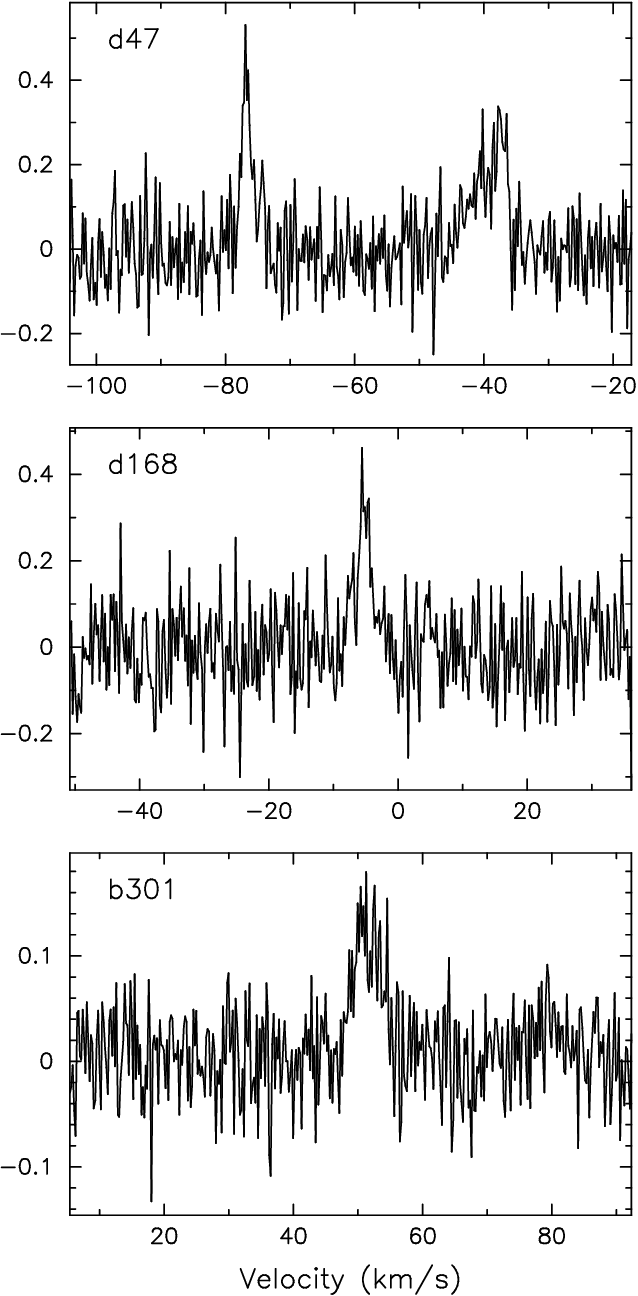}}
\figcaption{86 GHz SiO maser profiles for d47, d168 and b301 measured  
using the Mopra radio telescope on 2004 August 26.  The flux density scale
is uncalibrated. Velocities are accurate to $\pm$ 1 km s$^{-1}$. 
\label{Fi:sio_dets}}
\end{figure}

SiO masers at 86~GHz were detected for three sources: d47, d168 and
b301.  All three are LI sources, from the sample of six LI and five
non-LI sources observed.  Two of the objects (d47 and d168) were
detected on each of the observation dates with similar profiles each
time.  Figure \ref{Fi:sio_dets} shows the spectra from the average of
the two linear polarisations for the detections on 2004 August 26.

Discussion of the individual sources follows, including remarks on the
SiO maser detections, previous observations, unusual features and
likely blending of components.

\begin{deluxetable}{c c c c c c c c c c c}
\tablewidth{0pt}
\tablecaption{22~GHz H$_2$O maser results \label{Ta:water_results}}
\tablehead{
\colhead{Name} & \colhead{V$_{bp}$} & \colhead{S$_{bp}$} & \colhead{I$_b$} &  
\colhead{V$_{rp}$} & \colhead{S$_{rp}$} & \colhead{I$_r$} &  
\colhead{V$_{bl}$} & \colhead{V$_{rl}$}  & \colhead{Profile}  \\  
 & \colhead{{\ssize(km~s$^{-1}$)}} & \colhead{{\ssize(Jy)}} &  
\colhead{{\ssize(Jy~km~s$^{-1}$)}} & \colhead{{\ssize(km~s$^{-1}$)}} 
& \colhead{{\ssize(Jy)}} & \colhead{{\ssize(Jy~km~s$^{-1}$)}} &
\colhead{{\ssize(km~s$^{-1}$)}} & \colhead{{\ssize(km~s$^{-1}$)}}} 
\startdata
d46         & $-$121.0 &  9.3 &  66   &   \nodata & \nodata & \nodata & $-$145  & $-$54  & H  \\ 
(2)         & $-$136.9 &  9.3 &  62   &   \nodata & \nodata & \nodata & $-$145  & $-$34  & -- \\ 
d47         &  $-$41.4 &  6.3 &  37   &   \nodata & \nodata & \nodata &  $-$83  & $-$32  & I  \\ 
(2)         &  $-$41.4 &  5.0 &  34   &   \nodata & \nodata & \nodata &  $-$82  & $-$32  & -- \\ 
(3)         &  $-$38.3 &  2.3 &  11   &   \nodata & \nodata & \nodata &  $-$82  & $-$33  & -- \\ 
d62\tnm{a}  &  $-$73.6 &  0.9 &   0.5 &   \nodata & \nodata & \nodata &  $-$75  & $-$72  & H  \\ 
d103\tnm{a} &  $-$63.6 &  0.5 &   0.9 &   \nodata & \nodata & \nodata &  $-$65  & $-$62  & R  \\ 
d168        &  $-$21.7 & 12.1 &  30   &   \nodata & \nodata & \nodata &  $-$25  & $-$16  & R  \\ 
d189        &   $-$2.2 & 46.0 & 140   &   \nodata & \nodata & \nodata &   $-$5  &    16  & S  \\ 
d200        &  $-$87.4 &  0.6 &   1.1 & $-$59.4   &   0.4   &    0.9  &  $-$90  & $-$57  & R  \\ 
b25         &     14.9 &  0.6 &   0.3 &    21.2   &   1.0   &    1.5  &     11  &    22  & R  \\ 
b133        &  $-$16.8 &  0.3 &   0.4 &    15.7   &   1.3   &    4.4  &  $-$18  &    21  & R  \\ 
b292        &    100.2 & 25.3 &  51   &   \nodata & \nodata & \nodata &  $-$20  &   184  & H  \\ 
b301        &     26.4 &  1.9 &   6.1 &   \nodata & \nodata & \nodata &     14  &    43  & R  \\ 
v56         &   $-$1.7 &  0.5 &   1.1 &    32.5   &   0.9   &    2.3  &   $-$3  &    35  & R  \\ 
v121        &    101.9 &  0.3 &   1.1 &   \nodata & \nodata & \nodata &    100  &   107  & R  \\ 
v149\tnm{a} &     20.6 &  0.3 &   0.5 &   \nodata & \nodata & \nodata &     19  &    25  & R  \\ 
v154        &     29.0 &  9.3 &  50   &   \nodata & \nodata & \nodata &     23  &    66  & I  \\ 
v212        &     22.9 &  2.1 &   4.7 &    54.2   &   1.0   &    3.5  &     17  &    59  & R  \\ 
v223        &    122.7 & 60.2 & 190   &   \nodata & \nodata & \nodata &  $-$60  &   127  & H  \\ 
v228        &     85.0 &  0.4 &   0.7 &   \nodata & \nodata & \nodata &     83  &    90  & R  \\ 
v239        &     63.6 &  0.6 &   1.5 &   \nodata & \nodata & \nodata &     60  &    66  & R  \\ 
v268        &     32.1 &  7.6 &  60   &   \nodata & \nodata & \nodata &     29  &    74  & I  \\ 
v270        &     61.6 &  3.5 &   5.2 &   117.8   &   0.9   &    0.9  &     60  &   119  & H   
\enddata
\tnt{a}{Sources for which observations at two epochs were averaged together}
\end{deluxetable} 

\begin{deluxetable}{l c c c c c}
\tablewidth{0pt}
\tablecaption{Detection statistics for infrared 
groups and H$_2$O maser profile types
\label{Ta:water_stats}}
\tablehead{
\colhead{Frequency / Profile} & \colhead{Total} & \colhead{RI} & \colhead{LI} & 
\colhead{Quad I} & \colhead{Quad IV}} 
\startdata
\cutinhead{OH 1612 MHz maser detections}
1612 MHz & 85 & \ph{um}38\ph{um} & \ph{um}30\ph{um} & \ph{um}21\ph{um} & 15 \\  
\cutinhead{22 GHz H$_2$O maser detections by profile}
Total & 21 & 1 & 15 & 1 & 5 \\ Regular (R) & 12 & 0 & 11 & 0 & 1 \\ Single (S)
& 1 & 0 & 0 & 0 & 1 \\ Irregular (I) & 3 & 0 & 3 & 0 & 0 \\ High-velocity (H)
& 5 & 1 & 1 & 1 & 3
\enddata
\end{deluxetable}

\subsection{Individual Source Notes}   \label{Sec:water_source_notes}
 
This section gives notes for all sources with 22 GHz H$_2$O or 86 GHz
SiO detections from this study.  Notes are also given for sources that
were not detected but have been observed in previous 22 GHz H$_2$O or
43 or 86 GHz SiO studies.  Where the observing frequency is not
specified, the 22~GHz H$_2$O maser transition is implied. The central
velocity of a spectrum is defined as the velocity mid-way between the
most extreme blue and red-shifted emission at a detection level of
$\sim$ 3 sigma. 

{\it d3:} Not detected.  Observed at 86~GHz (no detection) by
\citet{Nym98}.

{\it d46:} This source has irregular and broad emission at the OH
(1612, 1665 and 1667) and 22~GHz H$_2$O transitions, with maser
features present over a velocity range of $-$200 to
$-$60~km~s$^{-1}$. The three OH profiles are blue-shifted
with respect to the H$_2$O maser profile, with an average of
50~km~s$^{-1}$ difference between the the OH and H$_2$O central 
velocities. The spectral profiles and peak and integrated intensities
at 22~GHz were similar on 2003 August 28 and November 15, although the
peak flux of the strong feature at a velocity of $-$121 km s$^{-1}$
decreased substantially between the two epochs.  A few weak 22~GHz
features detected just off the red-shifted end (up to
$-$34~km~s$^{-1}$) of the 2003 Nov spectrum are not plotted.
 
{\it d47:} This is the only source with a four-peaked `DD' OH 1612 MHz
spectral profile (Paper I).  22~GHz observations were taken on 2003
August 28, 2003 November 15 and 2004 May 20.  The peak intensity of the 
brightest features decreased between the first two epochs although the
integrated flux density over the spectrum remained constant.  By 2004
May most features had dimmed to about half their initial peak
intensity and the integrated flux had decreased by around 70 per cent. SiO
maser emission at 86~GHz was detected, with two features at
approximately $-$40 and $-$75~km~s$^{-1}$.
 
{\it d62:} The OH single-dish spectrum for d62 is cut at
$-$68~km~s$^{-1}$ where there is confusion with d59 (Paper I).  The
stellar velocity, determined from the OH 1612 MHz maser spectrum, is
$-118$ kms$^{-1}$ (Deacon et al. 2004). The single weak feature at
22~GHz is separated from the stellar velocity by about 45~km~s$^{-1}$,
and is well outside the OH 1612~MHz emission.  It is very close in
velocity to the OH emission from d59, but is probably not associated
with d59 as this source is more than 8~arcmin away, substantially
outside the Tidbinbilla beam.  No other 22~GHz sources are reported
near the stellar position and the source, though weak, was detected
twice on 2003 November 15 and 2004 May 20 (the average spectrum is
plotted). We infer that the H$_2$O maser emission is associated with
d62.
 
{\it d103:} This source was detected twice (2004 February 28 and May
20).  The averaged spectrum is plotted.
 
{\it d168:} Detected at both 22 GHz and 86 GHz.  The SiO maser feature
has a velocity of approximately $-$5~km~s$^{-1}$, similar to the
stellar velocity of $-4.5$ km s$^{-1}$ (Deacon et al. 2004).

{\it d189:} The only source with a narrow H$_2$O maser emission
profile centred on the stellar velocity determined from the OH 1612
MHz spectrum.
 
{\it d200:} Detected.

{\it b17:} Observed at 22~GHz (no detection) by \citet{Gom90}.

{\it b25}: Narrow double-peaked spectrum with peaks located $\sim$
7~km~s$^{-1}$ inside the 1612~MHz peaks. 

{\it b70:} Not detected.  Observed at 22~GHz (no detection) by
\citet{Lik89} and at 86~GHz (no detection) by \citet{Nym98}.  
 
{\it b96:} Not detected.  Observed at 22~GHz (no detection) by
\citet{Gom90}. 
 
{\it b128:} Not detected.  Observed at 86~GHz (no detection) by
\citet{Nym98}.

{\it b133:} Previously detected at 22~GHz by \citet{Tay93} with a
similar, but weaker ($\sim$0.8~Jy), double-peaked spectrum. 
 
{\it b134:} Not detected.  Observed at 22~GHz (no detection) by
\citet{Gom90} and at 86~GHz (no detection) by \citet{Nym98}. 
 
{\it b143:} Not detected.  Observed at 86~GHz (no detection) by
\citet{Nym98}.

{\it b251:} Not detected.  Observed at 22~GHz (no detection) by
\citet{Gom90}.
 
{\it b292:} This source has a very weak 1612~MHz peak near
100~km~s$^{-1}$ that is barely visible in Figure
\ref{Fi:H2O_thumbnails:c}.  Several blue- and red-shifted 22 GHz
features are detected at velocities that are well outside the velocity
range of the OH 1612~MHz emission (Figure \ref{Fi:b292_v223}).  A weak
red-shifted feature was detected at a velocity of 184 km s$^{-1}$ near
the edge of our bandpass and it is possible that other features may
extend beyond the range of velocities observed. No known H$_2$O maser
sources are located near this source within the beam. We infer that
the high-velocity features are associated with b292. This source is
also the only known post-AGB star with OH 1720~MHz maser emission
\citep{Sev01b}, and is discussed further in Section
\ref{Sec:new_water_fountain}.

{\it b301:} Detected.  Also detected at 22~GHz in 1987 by
\citet{Val01} but not detected by \citet{Tak01}. 86~GHz SiO maser
emission was detected at Mopra, with one feature at approximately
50~km~s$^{-1}$. 
  
{\it v45:} Not detected.  Observed at 22~GHz (no detection) by \citet{Gom90}.

{\it v56:} Detected.

{\it v67:} Not detected.  Detected at 22~GHz
\citep{Gom90,Ben96,Val01}.  In a 10 year monitoring program by
\citet{Eng02}, the H$_2$O emission was strongest in 1992 (4 Jy)
but faded and was undetected after 1997.  There was no detection by
\citet{Tak01}.  Detected at 43~GHz by \citet{Gom90} but not by
\citet{Nym98}.  Observed 86~GHz (no detection) by \citet{Nym98}.
 
{\it v87:} Not detected.  Detected at 22~GHz (1990) by \citet{Val01}
but no detection in \citet{Tak94}.
 
{\it v117:} Not detected.  Observed at 22~GHz (no detection) by \citet{Gom90}.

{\it v121:} Marginal detection.  Not detected by \citet{Tak01}.
Detected at SiO 43~GHz with no 86~GHz detection \citep{Gom90,Nym98}.
 
{\it v132:} Not detected.  Detected at 22~GHz by \citet{Nym86} and
\citet{Lik89}; this source was a $\ge 50$~Jy H$_2$O maser source in
the 1980's.  It was monitored for 10~years by \citet{Eng02}, growing
weaker and finally disappearing after 1990 \citep{Eng02,Tak01}.
Observed at 86~GHz (no detection) by \citet{Nym98}
 
{\it v149:} Observed twice, on 2004 February 28 and May 20. The
averaged spectrum is plotted.
 
{\it v154:} The blue-shifted peak was detected by \citet{Gom90},
\citet{Tak94}, \citet{Tak01} and \citet{Val01}, with the most
red-shifted peak faint or undetected. Detected at 43~GHz
\citep{Gom90}. 
 
{\it v162:} Not detected.  Observed at 22~GHz (no detection) by 
\citet{Ben96}.   

{\it v169:} Not detected.  Observed at 22~GHz (no detection) by 
\citet{Gom90}. 

{\it v172:} Not detected.  Observed at 22~GHz (no detection) by 
\citet{Gom90}. 

{\it v204:} Not detected.  Observed at 22~GHz (no detection) by 
\citet{Gom90}. 

{\it v211:} Not detected.  Detected at 22 and 43~GHz by \citet{Gom90}.
Observed at 22~GHz (no detection) by \citet{Nym86}. 

{\it v212:} Detected.

{\it v223:} Extreme blue- and red-shifted features are present.  This
star has been imaged previously at OH (1612~MHz), H$_2$O (22~GHz) and
SiO (43~GHz) maser frequencies \citep{Ima02,Ima05} confirming the
object as a post-AGB source with high-velocity H$_2$O jets located
outside the shell of OH masers.  The H$_2$O maser spectrum is highly
variable \citep{Lik92}. Observed at 86~GHz (no detection) by
\citet{Nym98}.  Discussed further in Section
\ref{Sec:v223_water_disc}.

{\it v228:} Detected.  Also detected at 22~GHz by \citet{Gom90} and
\citet{Val01} and detected at SiO 43~GHz by \citet{Gom90} and \citet{Nym98} 
(no detection at 86 GHz).
 
{\it v231:} Not detected.  Observed at 22~GHz (no detection) by 
\citet{Gom90}. 

{\it v239:} Detected.

{\it v268:} About eight times stronger at 22~GHz than at 1612~MHz.  
 
{\it v270:} A double-peaked profile with each peak 15~km~s$^{-1}$
outside the 1612~MHz maser peaks.  Previously detected at 22~GHz by
\citet{Eng02}, \citet{Gom94}, \citet{Tak01} and \citet{Val01}.  The 
fainter red-shifted peak is not always detected \citep[e.g.][]{Gom94},
but there have been no other significant changes in the spectrum.
Observed at 86~GHz (no detection) by \citet{Nym98}. Discussed further
in Section \ref{Sec:new_water_fountain}.
 
{\it v274:} Not detected.  Observed at 22~GHz (no detection) by 
\citet{Lik89}.
 
\section{SAMPLE ANALYSIS} \label{Sec:water_discussion}
 
\subsection{Profile Types and IR colours} \label{Sec:water_profile_types}

Of the 21 H$_2$O detections, 16 have available IRAS colours: 15 of
these are LI sources and one is an RI source. Six sources have MSX
colours with five identified as Quad IV (young post-AGB) and one as
Quad I (older post-AGB).

Twelve sources have regular (R) H$_2$O maser spectra (Tables
\ref{Ta:water_results} and \ref{Ta:water_stats}), with velocity widths
equal to or slightly smaller (by 5 to 10~km~s$^{-1}$) than their OH
spectra, and either single- or double-peaked profiles.  Single-peaked
regular spectra have features at velocities located near the outer
blue- or red-shifted OH emission peaks.  The profiles of these sources
indicate that the H$_2$O masers are located in regions of their
circumstellar envelopes where the velocity gradients are small and
radial amplification dominates, as for the OH 1612~MHz masers in many
OH/IR sources. As noted in previous studies, blue-shifted features
tend to be brighter than the red-shifted features. Of the 12 sources
with regular profiles, eight have strongest emission at blue-shifted
velocities while four have strongest emission at red-shifted
velocities.  

Five high-velocity sources (d46, b292, v223, d62 and v270) were
detected. For each of these the velocity range of the H$_2$O maser
emission is broader than that of the OH. The detection of these
sources provides strong evidence for a `turn-on' of high-velocity
outflows in the lower mass post-AGB stars. It is remarkable that for
four of the five sources the H$_2$O maser emission is strongest on the
{\it red-shifted} side, in contrast to the sources with regular
profiles. We are unable to explain this difference. The high-velocity 
H$_2$O maser sources are discussed further in 
Section \ref{Sec:high_velocity_water}.

Three LI sources (d47, v154 and v268) have irregular spectra with many
emission features over their profiles.  These sources clearly have
complex masing regions, probably with a combination of tangential and
radial amplification. For each of these the velocity range of the
H$_2$O maser emission lies within that of the OH maser emission. 

For 16 of the 21 candidate post-AGB stars, the peak H$_2$O maser flux
density is weaker than the peak OH 1612 MHz maser flux density. Five
sources (d46, b292, v154, v223 and v268) have stronger H$_2$O
masers. These include three of the five high-velocity sources (d46,
b292 and v223) and two LI sources with irregular spectra (v154 and
v268). 

Only one source, d189, has an S profile, with several closely-spaced
features near the stellar velocity.  This spectrum is similar to that
of some Miras where the H$_2$O masers are located in a inner,
accelerating region and so are tangentially amplified.  With a red
IRAS [12--25] colour of 1.2, it is unusual for d189 to have such a
narrow Mira-like profile (Section \ref{introduction}). There is also
evidence for an accelerating outflow in the OH mainline spectra of
d189.  A central plateau of emission is present in both the 1665 and
1667~MHz spectra (Paper I).  Both mainline spectra also have features
at velocities near the 1612 MHz peaks that probably arise from a
remnant regular circumstellar envelope.  The plateau mainline emission
covers a similar velocity range to the H$_2$O maser emission.

There is no evidence in this study for an evolution from R to S
profiles at IRAS [12--25] $\ge$ 0.5 as suggested by \citet{Eng02}.  In
that work the detection of Mira-like S profiles from some post-AGB
stars was considered to be due to decreasing mass-loss rates as stars
evolved away from the AGB. 

\begin{figure}
\centering
\rotatebox{270}{\scalebox{0.3}{\plotone{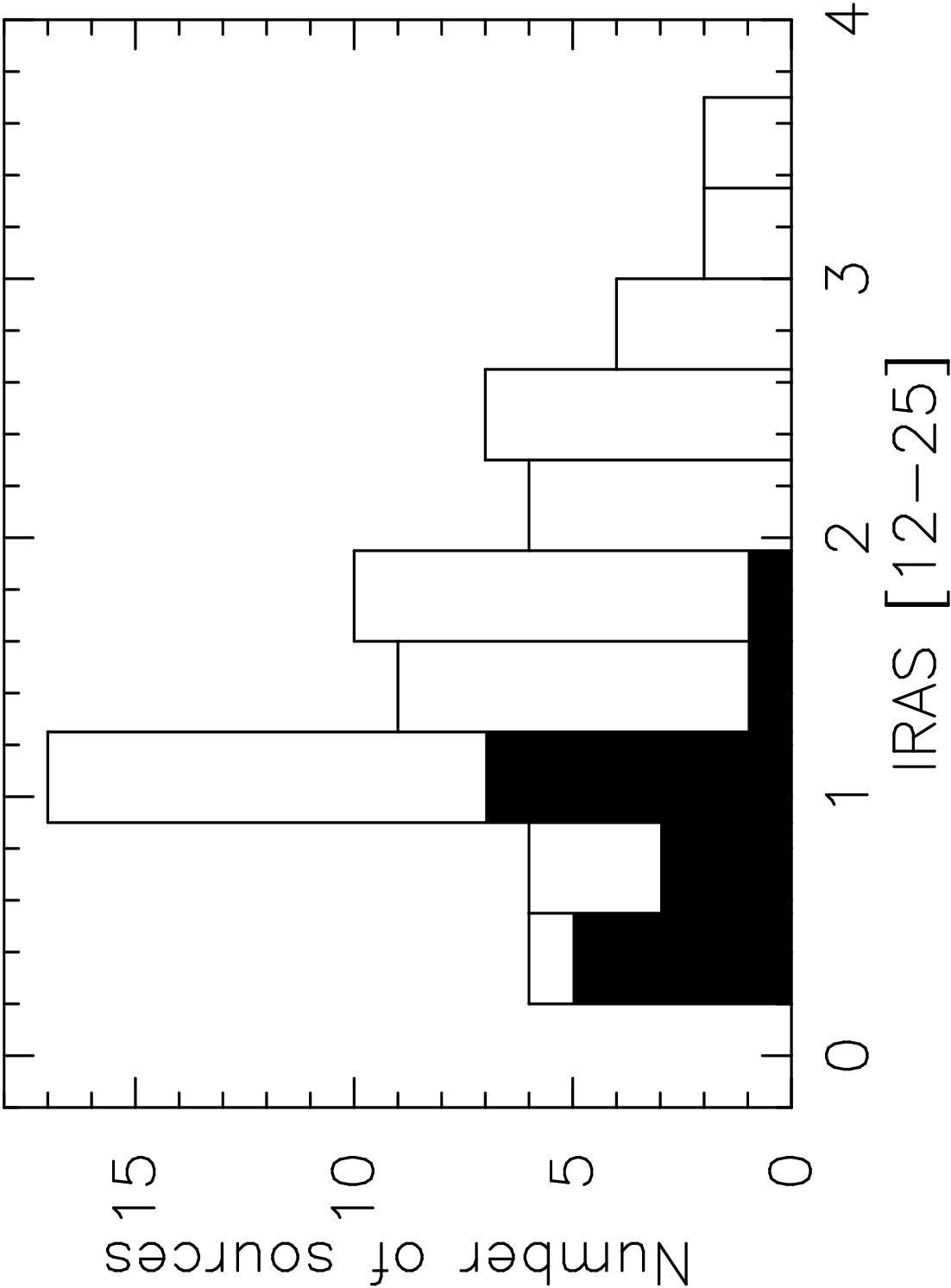}}}
\hspace{5pt}
\rotatebox{270}{\scalebox{0.3}{\plotone{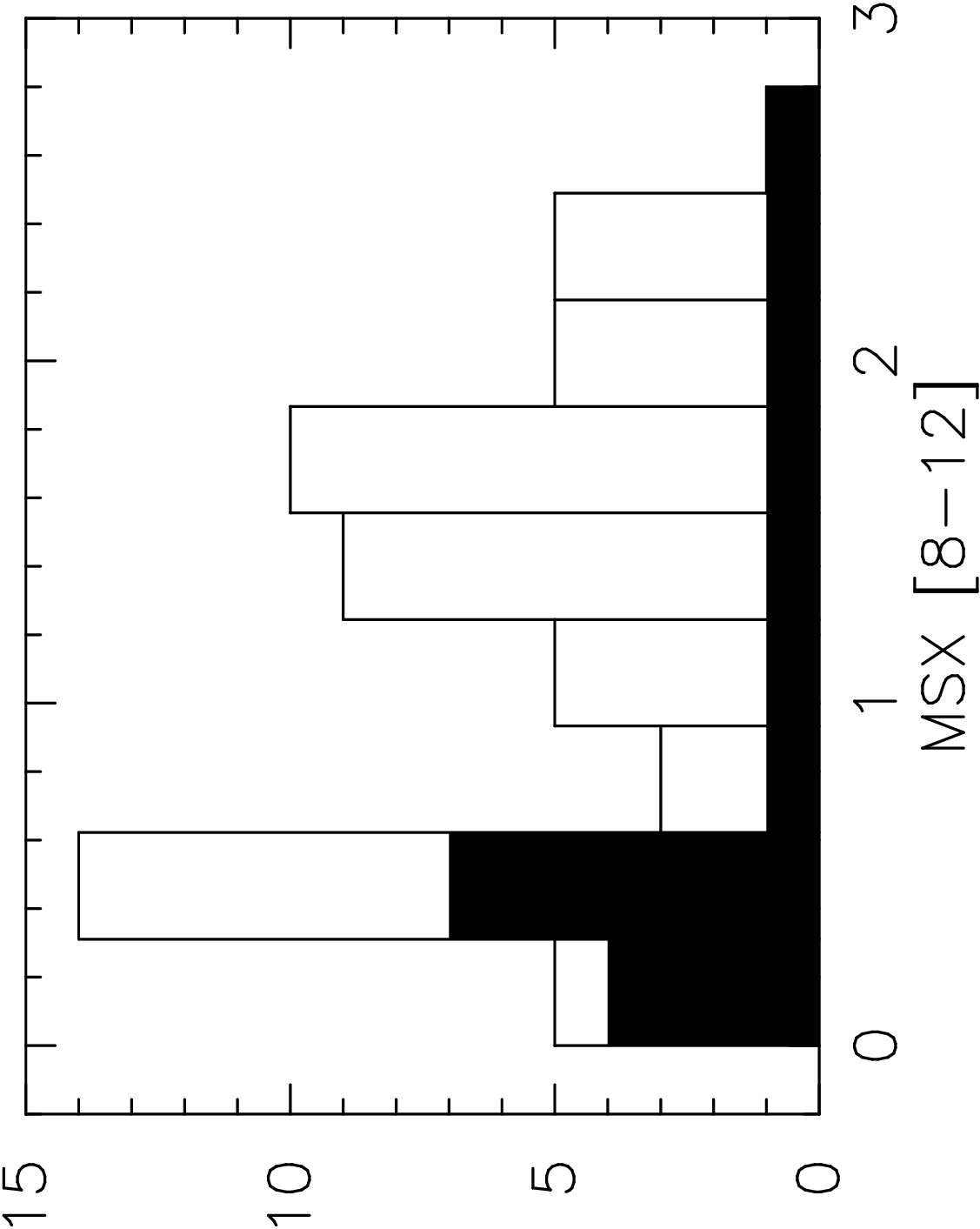}}}
\figcaption{Two histograms of the number of sources observed (outline) 
and detected (filled) at 22~GHz versus IRAS [12--25] ({\it left}) and
MSX [8--12] ({\it right}) colours.  The overall detection rates were
25 per cent for the 69 sources with IRAS colours, and 32 per cent for
the 57 sources with MSX colours. \label{Fi:water_dets_12-25_8-12}}
\end{figure}

Figure \ref{Fi:water_dets_12-25_8-12} gives histograms for the source
detections for IRAS [12--25] and MSX [8--12] colours, showing the drop
in H$_2$O detection rates at redder colours.  For 69 sources with
available IRAS [12--25] colours, the detection rate is $\sim$ 50 per
cent for [12--25] $<$ 1.3. There is a clear drop in the detection rate
at [12--25] $\sim$ 1.3, with only 2/40 sources detected at redder
colours, and no detections for [12--25] $>$ 1.9.  Similar results were
found by \citet{Eng96} and \citet{Tak01}. For 57 sources with MSX
colours, the detection rate is $\sim$ 55 per cent for [8--12] $<$ 0.6
with a sharp drop in the detection rate at [8--12] $\sim$ 0.6.
However, a larger fraction of sources (7/38) were detected at redder
colours.  Five of these are high-velocity sources.

From our sample of 85 sources we find that the detection rates are
consistent with a strong decrease in the H$_2$O maser emission from
the circumstellar envelope as a star leaves the AGB.  However, in some
sources, a new source of strong H$_2$O maser emission turns on in the
post-AGB phase with high-velocity outflows and/or irregular
profiles. As discussed in Section 6, the detection of such H$_2$O
maser spectra is likely, in at least some cases, to be associated with
the emergence of collimated jets from the central star and the onset
of morphological changes in the circumstellar envelopes as a star
evolves away from the AGB.

\subsection{H$_{\mathbf{2}}$O and SiO Masers in LI Sources} \label{Sec:LI_are_AGB}

Of the 21 sources detected, 15 are LI sources giving a 50 per cent
detection rate for this subset. The LI sources have very different OH
and H$_2$O maser characteristics to the rest of the sample.  As
discussed in Paper I, their OH 1612 MHz spectral profiles are almost
all double peaked with higher than average envelope expansion
velocities. The lack of OH mainline masers (Paper I) and the high
detection rate (50 per cent) for H$_2$O maser emission indicates
cooler, denser envelopes than for most other sources in the sample.

The OH and H$_2$O maser properties of LI sources are consistent with a
classification as high-mass AGB stars with high mass-loss rates.  In the inner
envelopes, the gas densities are more favourable for H$_2$O masers
than OH mainline emission while at larger radii the photo-dissociation
of H$_2$O molecules leads to OH 1612 MHz maser emission. The
double-peaked OH 1612 MHz spectral profiles show that the envelopes
are still mostly spherical.

Few conclusions can be drawn from the small number of 86 GHz SiO maser
observations conducted.  However, the detection of SiO maser emission from 
three LI sources is consistent with an evolutionary status as AGB stars.
Most of the 43~GHz SiO maser detections of this sample from previous
studies are also LI sources (Section \ref{Sec:water_source_notes}; the
exception is v223).

The wide double-peaked SiO maser profile found in d47 is unusual as very
few SiO maser sources have double-peaked profiles.  This sources also
has an unusual four-peaked 1612~MHz OH maser profile and biconal
outflow \citep[Paper I,][]{Cha88}.  Several geometries may explain the
SiO maser profile: a very large circumstellar envelope from a massive
star as in IRAS 19312$+$1950 (Section \ref{Sec:sio_late_type}); a
torus as in OH 231.8$+$4.2, or a biconal SiO maser outflow as in v223
(W43A) \citep{Ima05}.  Further observations of this unique source are
needed to investigate these possibilities.

\subsection{H$_{\mathbf{2}}$O Maser Variability}

Multiple observations of some of the sources enabled the variability
of H$_2$O maser features to be assessed.  Spectral features are stable
in velocity over a time-scale of a few months, as shown by the profile
constancy in the multiple observations of d46, d47 and the repeat
detections of d62, d103 and v149.  However, the peak intensity of
individual features and the integrated spectral flux density vary
dramatically.  A lifetime for individual H$_2$O masers in evolved
stars of 1--3~years \citep[e.g.][]{Eli92a,Bai03b}, with high variability
and low or no saturation, is consistent with these results.

\clearpage

\section{DISCUSSION}
\subsection{High-Velocity and Water-Fountain H$_{\mathbf{2}}$O Masers}   
\label{Sec:high_velocity_water}

We detected five sources with H$_2$O masers at velocities well outside
their OH spectral range. Identifications for these sources are: v223
(W43A, IRAS18450$-$0148, OH 31.0$+$0.0), v270 (IRAS 18596$+$0315, OH
37.1$-$0.8), d46 (IRAS 15445$-$5449, OH 326.5$-$0.4), d62 (IRAS
15544-5332, OH 328.5$-$0.3) and b292 (IRAS 18043$-$2116, OH
0.9$-$0.4). Three sources, b292, v223 and v270, have MSX
classifications as Quad IV sources (young post-AGB), one source, d46,
is a MSX Quad I source (older post-AGB) while d62 is an IRAS LI
source.

Two of the five sources, v223 and v270, have previously known
H$_2$O maser emission. The source v223 (W43A) is well-known and belongs to 
a rare group of sources first described by \citet{Lik88} as `water-fountains'.
 These are characterised by H$_2$O maser emission over an usually large 
velocity range. High resolution observations have 
revealed well-collimated bipolar jets. The H$_2$O masers are probably 
excited in post-shock regions as the expanding jets hit the ambient
circumstellar material. In some cases the H$_2$O masers exhibit a
strong degree of symmetry with pairs or groups of blue-and red-shifted
emission features symmetrically placed on either side of the stellar
position. The dynamical ages estimated for the jets are very short,
typically around 100 years.

To date only four sources, v223 (W43A), IRAS 16342$-$3814 (OH
344.1$+$5.8), IRAS 19134$+$2131 and OH 12.8$-$0.9, are confirmed
members of this class. The source v270 is also a likely water-fountain
source although this has not yet been confirmed from high angular
resolution observations. The properties of these five sources are
briefly reviewed below:

\paragraph{v223 (W43A, IRAS 18450$-$0148, OH 31.0$+$0.0)} 
\label{Sec:v223_water_disc}

Our spectrum of W43A from May 2004, plotted in Figures
\ref{Fi:H2O_thumbnails:d} and \ref{Fi:b292_v223}, shows strongest
emission in two velocity ranges near 120 and --60 km s$^{-1}$, with
strongest emission of 60 Jy at a red-shifted velocity of 122.7 km
s$^{-1}$. Likkel et al. (1992) previously monitored the H$_2$O
emission from 1987 to 1989. A comparison of their spectra shows the
strongest emission over the same velocity ranges although the
velocities of the strongest features are not the same. Strong emission
at similar velocities was detected earlier by Genzel \& Downes (1977)
and Diamond et al. (1985). It thus appears that the velocity range of
the stronger emission has remained stable over almost 30 years,
although the individual emission features vary strongly. We note that
the ratio of the strongest red-shifted emission to strongest
blue-shifted emission was much larger in 2004 (10:1) than in the
spectra taken approximately 15 years earlier (2:1). The Tidbinbilla
spectrum also shows some additional weak features near --30, 0 and 80
km s$^{-1}$ while the Likkel et al.  data show additional features
near 7 and 19 km s$^{-1}$ that were not detected in 2004.

The high-velocity H$_2$O maser emission from v223 (W43A) occurs from
two narrow jets on opposite sides of the central star, twice as far
out as the OH masers \citep{Ima02, Ima05} while emission at velocities
close to the stellar velocity, is located inside the OH
masers. Between 1994 and 2002 the separation between the most blue-
and red-shifted H$_2$O maser clumps increased from 1700 to 2400 AU.  A
precessing jet model for the H$_2$O jets gives a dynamical age of only
$\sim 50$ years, a precession period of 55 years, and a
three-dimensional speed of 145~km~s$^{-1}$ at an inclination of
39\degr\ with respect to the plane of the sky \citep{Ima05}.

The H$_2$O masers in W43A are aligned along the jet axis.  Recent
polarisation observations have estimated the magnetic field strength
in the jet to be $200 \pm 75$ milligauss (mG) and a toroidal
configuration such that the field varies as $r^{-1}$ where $r$ is the
radius from the star \citep{Vle06a}.  Extrapolating back to the
stellar surface, this gives an equatorial magnetic field of 35 gauss
(G) (if the jet H$_2$O masers are created in compressed, swept-up
material) or 1.6 G (if the jet H$_2$O masers are created in shocks).
The magnetic field strength and configuration suggests that the jet in
v223 is magnetically collimated and is the most direct evidence thus
far for magnetic collimation in evolved stars.

The OH 1612 MHz maser emission from W43A is double peaked with two
clumps of masers separated on the sky and an OH expansion velocity of
approximately 10 km s$^{-1}$. The blue-shifted OH masers and
red-shifted H$_2$O masers are located on the NE side of the star and
the red-shifted OH and blue-shifted H$_2$O masers on the SW side.
This may be explained if the OH masers are located in an equatorial
plane perpendicular to the H$_2$O jet. The dynamical age of the OH
masers is $\sim$ 2600 years \citep{Ima02}.

W43A also has remarkable 43 GHz SiO maser emission. The SiO masers
were imaged with the VLA and show a biconical outflow that is parallel
to the H$_2$O jet, lying inside the OH masers (Imai et al. 2005).
The SiO masers could be excited by shocks in the region between the jet
and the OH masers.

\paragraph{v270 (IRAS 18596$+$0315, OH 37.1$-$0.8)}

H$_2$O maser emission from v270 was first detected in 1984 
\citep{Eng86}, but was initially considered to be unrelated because the
velocity was outside the range of the OH masers.  Our 2004 May H$_2$O
maser spectrum of v270, shown in Figure \ref{Fi:H2O_thumbnails:d},
shows two narrow regions of emission, each approximately 30 km
s$^{-1}$ from the stellar velocity, approximately twice the outflow
velocity of the OH maser emission.  Engels (2002) monitored the 22 GHz
emission from v270 between 1990 and 1999 and identified nine spectral
components with lifetimes for the individual components of typically
2--3 years.  Our more recent spectrum shows the same overall spectral
structure although the individual components have changed.

VLA observations of v270 by \citet{Gom94} confirmed the coincidence of
the OH and H$_2$O maser positions to within 1 arcsec. Later VLBA
observations of the OH 1612~MHz masers found a bipolar morphology. The
source is $\sim8$ kpc away \citep{Bau85} and the red- and
blue-shifted masers are in clumps that are separated by $\sim1300$ AU
\citep{Gom00} probably indicating a bipolar outflow.  The H$_2$O
masers have not been imaged. From the symmetry of the source and the
high-velocity H$_2$O features, we consider v270 to be a likely
water-fountain source.

\paragraph{IRAS 16342$-$3814 (OH 344.1+5.8)}

This outflow source was first discovered by \citet{Lik88} and
\citet{teL88} and shows high-velocity OH and H$_2$O masers. Hubble 
Space Telescope (HST) images show two asymmetric bipolar reflection
lobes on either side of a dark lane. The OH masers are aligned with
the inner edges of the lobes, with outflow velocities up to 70 km
s$^{-1}$ \citep{Sah99,Zij01}. Near-infrared images \citep{Sah05a}
reveal corkscrew-like structures in the lobes which are consistent
with a collimated precessing jet of diameter $\leq$ 100 AU and period
$\leq~ 50$~years.

H$_2$O maser emission from IRAS 16342$-$3814 is detected at radial
velocities up to 160 km s$^{-1}$ from the stellar velocity
\citep{Lik88,Lik92,Mor03}. High angular resolution observations with
the Very Long Baseline Array \citep[VLBA,][]{Mor03,Cla04} have shown
that these masers are located near the outer edges of the optical
lobes close to the polar axis of the nebula.  The red- and
blue-shifted groups of H$_2$O maser spots show quasi-linear formations
perpendicular to the nebula axis, consistent with bow shocks.  For an
estimated distance of 2 kpc this corresponds to a separation of 6000
AU, although \citet{Zij01} used a closer distance of 700 pc.  The
dynamical age of the jets is estimated to be 150 years \citep{Cla04}.

\paragraph{IRAS 19134$+$2131}

This has two groups of H$_2$O maser features, separated by 100
km~s$^{-1}$.  There have been no detections of OH maser
emission \citep{Lik92}.  The H$_2$O maser emission also reveal
bipolar jets, with a dynamical age of around 50 years old, although
these are less collimated than for W43A \citep{Ima04}.  From
measurements of the secular motion of the maser spots, a far kinematic
distance of $\geq$ 16 kpc has been determined for this source
\citep{Ima04}.
 
\paragraph{OH 12.8$-$0.9}

This is the most recently confirmed water-fountain source
\citep{Bob05}.  It has a regular double-peaked OH 1612 MHz maser
profile (24~km~s$^{-1}$ width) typical of an OH/IR star, and a H$_2$O
maser profile that is also double-peaked but with a velocity width
about twice as large (48~km~s$^{-1}$). Similarly for IRAS
16342$-$3814, the H$_2$O masers occur in arcuate structures at the end
of bipolar jets, with the blue- and red-shifted groups of masers
separated by $\sim 110$ mas.  The jet opening angle is estimated to be
10\degr--13\degr\ \citep{Bob05}.  The distance to OH 12.8$-$0.9 is
uncertain, but assuming the source is close to the Galactic Centre
with a distance of 8 kpc, the spatial separation on the sky of the
clumps is about 870 AU.  This gives an upper limit of 110 years on the
dynamical age of the jets.  The source has a lower outflow velocity
and smaller jet length than other water-fountain sources, but its
likely AGB/post-AGB status combined with the H$_2$O jet, firmly
classify it as such.

\subsection{A New Water-Fountain Source: b292} \label{Sec:new_water_fountain}

\paragraph{b292 (IRAS 18043$-$2116, OH 0.9$-$0.4)} 

We report the discovery of a new water-fountain source, b292.  H$_2$O
maser emission from b292 was first detected from a `snapshot'
observation taken with the ATCA in 2002 (see Paper I). The velocity
coverage of the 2002 observation was limited to velocities between
approximately 65 and 110 km s$^{-1}$. Three maser features were
detected at velocities of 73, 85 and 106 km s$^{-1}$.

Figure \ref{Fi:b292_v223} shows the more recent spectrum, obtained in
2004 March with the Tidbinbilla telescope, with the much wider velocity
coverage of approximately 210 km s$^{-1}$. This shows the discovery of
many discrete features over velocities between $-$20 and 185 km
s$^{-1}$. It is possible that even higher velocity features may exist
outside this velocity range. The H$_2$O maser emission was detected
within the 50 arcsec beam of the Tidbinbilla telescope and is
almost certainly associated with the evolved star, which has a systemic 
velocity of 87 km s$^{-1}$.

A comparison of the ATCA and Tidbinbilla spectra for the
inner velocity range shows that the maser emission is highly variable
and may be rapidly increasing in strength. In the more recent
spectrum, the H$_2$O maser emission was strongest at velocities
between $\sim$ 95 and 120 km s$^{-1}$ corresponding to the red-shifted
side (see below). The strongest emission peak has a flux density of 25
Jy at a velocity of 100 km s$^{-1}$. This feature was not detected
in the spectrum obtained just 18 months earlier. No obvious pair
symmetry is seen in the higher velocity extreme red- and blue-shifted
features.

The OH maser properties of b292 have been previously discussed by
\citet{Sev01b} and in Paper I. The source is unusual in having emission at
OH 1612 and 1665 MHz, but not at 1667 MHz. Both the 1612 and
1665 MHz emission extend over $\sim$ 33 km s$^{-1}$. It is also the only
post-AGB star with detected OH 1720 MHz maser emission, although
1720 MHz emission has been recently detected in the PN K3--35
\citep{Gom05b}, also an H$_2$O maser source.  From
\citet{Sev01b}, 1720 MHz masers are produced behind a C-type
shock under very specific physical conditions.  The density increases behind 
the shock front, leading to the formation of H$_2$O molecules. Subsequent 
dissociation of the H$_2$O can then lead to the shock-excited OH emission at 
1720 MHz. \citet{Sev01b} predicted that no H$_2$O maser emission
would be detected from b292 as the density behind the shock front
would not increase sufficiently for H$_2$O maser emission to
occur. However, the new detection of high-velocity H$_2$O emission
from b292 and the detection of OH 1720 MHz emission in K3--35
provides strong evidence that sufficiently high densities can occur.

Although no high resolution images have been obtained, it appears
likely that both the OH 1720 MHz maser emission and the higher
velocity H$_2$O maser features are associated with a high-velocity
jet, with the H$_2$O maser emission located at greater distances.
The OH 1612 and 1665 MHz masers are most likely
associated with the remnant circumstellar envelope. It is unclear
whether some H$_2$O maser features with velocities near the stellar
velocity may still be present inside the OH radius.

From the extremely broad H$_2$O spectrum we identify b292 as highly
likely to be a water-fountain source and note that the range of H$_2$O 
velocities, compared with the OH velocity range, is the most extreme known. 
High angular resolution and monitoring observations of this source would be 
valuable in confirming the proposed model.

\subsection{Two High-Velocity Sources: d62 and d46}

\paragraph{d62 (IRAS 15544$-$5332, OH328.5$-$0.3)}

This is classified as an LI source and so is likely to be from
a massive progenitor star. The irregular OH spectrum is unusual for an LI 
source with 1612, 1665 and 1667 MHz emission detected at
velocities between $-$100 and $-$130 km s$^{-1}$ and the strongest
integrated emission measured at OH 1667 MHz (Paper I). The OH spectra indicate
a stellar velocity near $-$115 km s$^{-1}$ although the velocity
is not well determined.

The 22 GHz spectrum of d62, obtained from observations in November
2003 and May 2004, shows a single emission feature at a velocity of
$-$74 kms $^{-1}$, offset by $\sim$ 40 km s$^{-1}$ from the central OH
velocity (Figure \ref{Fi:H2O_thumbnails:a}). A corresponding
blue-shifted peak (expected near $-$155 km s$^{-1}$) has not been
detected. The single high-velocity red-shifted emission feature
suggests it is possible that d62 is a water-fountain source, although
further observations are needed to clarify the classification.

\paragraph{d46 (IRAS 15445$-$5449, OH 326.5$-$0.4)}

\begin{figure}
\centering
\scalebox{0.8}{\plotone{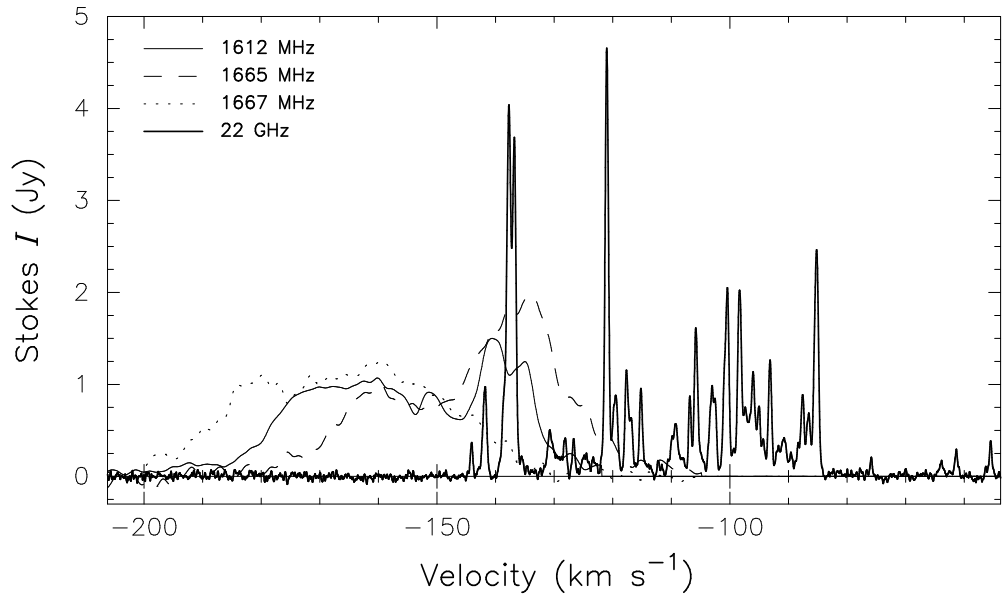}}
\figcaption{The OH 1612, 1665, 1667 MHz and 22 GHz H$_2$O maser emission 
from d46, with profiles plotted using the key given. The OH data are
from Paper I but Gaussian-smoothed for clarity.  The H$_2$O maser
emission is from the 2003 August observation and is scaled by 0.5.
\label{Fi:d46_water_oh}}
\end{figure}

The source d46 is highly unusual. It was first detected in 1994, in
the OH transitions at 1612 MHz \citep{Sev97b} and 1665 MHz
\citep{Cas98}, and in 1998 at 1667 MHz (Sevenster \& Chapman 2006, private
communication). The OH profiles are all extremely wide, with emission
velocities between $-$200 and $-$100 km s$^{-1}$. However, the OH 1667
MHz spectrum is blue-shifted relative to the OH 1612 MHz spectrum,
while the OH 1665 MHz spectrum is slightly red-shifted.  Figure
\ref{Fi:d46_water_oh} displays the OH emission from the three
frequencies 1612, 1665 and 1667 MHz, and the 22 GHz H$_2$O maser
emission, overlaid to show the offsets between the lines. The H$_2$O
profile is red-shifted by approximately 50 km s$^{-1}$ relative to the
OH profiles. The shape and width of the OH profiles are somewhat
similar to those of some bipolar post-AGB high-velocity sources,
including IRAS 18491$-$0207 and IRAS 11385$-$5517
\citep[e.g.][]{Zij01}.  However, the velocity shifts between the lines
are exceptional.  A comparison of several peak flux densities between
1994 and 2003 indicates that the OH maser emission is increasing in
strength, with peak 1612 MHz flux densities of 0.8 Jy in 1994, 1.2 Jy
in 1998 and 1.8 Jy in 2003.

Figures \ref{Fi:H2O_thumbnails:a} and \ref{Fi:d46_water_oh} show the
OH 1612 MHz and H$_2$O spectra for d46. The H$_2$O spectrum has
numerous narrow features blended together at velocities between $-$145
and $-$54 km s$^{-1}$. The two spectra taken in August and November
2003 show essentially the same spectral features, although with
different peak flux densities. Given the different velocity ranges of
the maser profiles, it is difficult to estimate the stellar
velocity. However, it seems plausible to take a value close to the
mid-point of the OH profiles, near a velocity of $-$150 km
s$^{-1}$. This would mean that the high-velocity H$_2$O masers have
been produced at the back of the circumstellar envelope, perhaps from
a one-sided jet-like structure.

From the radio continuum observations taken in November 1998, radio
continuum was detected from d46 with Gaussian-fitted flux densities at
3, 6 and 13 cm of 11, 18 and 30 mJy respectively. These correspond to
a steep non-thermal spectrum with a spectral index of $-$0.8. With the
limited {\it u-v} coverage of the ATCA observations, it was not
possible to determine if the emission was spatially resolved.

As discussed by \citet{Coh06}, nonthermal radio continuum emission
with a spectral index of $-$0.8 is typical of Wolf-Rayet (WR) stars in
binary systems, where synchroton radio continuum emission occurs from
the shocked regions at the interaction between the winds from two
massive stars. For WR stars, the 6-cm radio continuum luminosity is
typically 2 x 10$^{19}$ erg s$^{-1}$. Nonthermal radio continuum has
also been discovered from the PN V1018 Sco with a 6-cm radio continuum
luminosity of 4.4 x 10$^{19}$ erg s$^{-1}$ (Cohen et al. 2006). This
remarkable source is a near-LI star (Figure \ref{Fi:IR_water}) where
an optical PN is seen around a still pulsating AGB star (Cohen et
al. 2006). The radio luminosity in V1018 Sco is comparable to that of
WR stars but the synchroton emission is associated with the interface
between a hot fast wind and the slow AGB wind. For d46, a kinematic
distance of 7 kpc \citep[calculated assuming a flat Galactic rotation
model, a circular rotation speed of 220 km s$^{-1}$ and a Galactic
Centre distance of 8.5 kpc; e.g.][]{Fis03} combined with the 6-cm flux
density of 18 mJy, produces an estimate for the 6-cm luminosity of
$10^{21}$ erg s$^{-1}$, more than an order of magnitude higher than
for V1018 Sco and higher than for most WR stars. For an assumed
bandwidth of 10 GHz, the total radio luminosity is $\sim$ 10$^{31}$
erg s$^{-1}$. By comparison with WR stars where the ratio of kinetic
energy to radio luminosity is about 10$^{6}$:1, this indicates an
available kinetic energy, in the shock regions, of approximately
10$^{37}$ erg s$^{-1}$.

The presence of high-velocity OH and H$_2$O masers and non-thermal
radio continuum emission, together with the MSX classification of d46
as an older post-AGB star, are all consistent with its status as an
evolved post-AGB star with a highly disturbed envelope, where a fast
wind has ploughed into the slow AGB wind creating strong shocks and
generating synchroton radio continuum and the H$_2$O masers. High
resolution observations are needed to determine the structure of this
extraordinary source.

\subsection{Jet Origins}

As discussed above, there are now eight high-velocity H$_2$O maser
sources associated with evolved stars. For seven of them OH maser
emission is also detected. From the OH and H$_2$O maser properties and
their far infrared colours, four sources (W43A, IRAS 19134+2134, v270
and b292) are likely to be younger post-AGB stars. Two sources (IRAS
16342-3814 and d46) have extremely red far-infrared colours and broad
OH emission and are older post-AGB stars, and two sources (d62 and
OH12.8$-$0.9) are probably high-mass AGB stars. The observations are
consistent with a scenario where a high-velocity wind is generated as
a star leaves the AGB. The detection of two high-velocity sources in
the LI stars suggests that for some massive stars the jets may turn on
before the star has left the AGB. Initially the narrow collimated jets
pierce through the circumstellar envelope, which is still largely
spherical and may still have the classical double-peaked OH 1612 MHz
spectral profiles.  At a later stage the narrow jets may disappear but
the bipolar lobes will become wider as the fast and slow winds
continue to interact. In the `late' post-AGB phase the remnant
circumstellar envelope may appear highly disrupted with high-velocity
bipolar flows also seen in OH.

The origin of high-velocity collimated jets in post-AGB stars is not
yet fully understood. The `Generalised Interacting Stellar Winds'
model \citep{Kwo78,Bal87}, where a fast wind is shaped by a
pre-existing asymmetry in the slow wind, can explain a broad range of
envelope asymmetries, including bipolar and elliptical
outflows. However, it does not provide an explanation for the high
wind momentum, the highly-collimated jets seen in the water-fountain
sources, or the point and multi-polar symmetry of some PN
\citep[e.g.][]{Buj01,Fra04}.

Recent observations and theoretical models suggest that the collimated 
jets are shaped by stellar magnetic fields. From VLBA observations 
Vlemmings et al. (2006) have measured
strong linear and circular polarisation in the H$_2$O maser features
of W43A, at distances up to 1000 AU from the central star. From the
estimated field strengths they argue that the magnetic field is
collimating the jet. \citet{Gre02} has detected a toroidal magnetic
field around the PN NGC 7027.

Theoretical models show that strong magnetic fields may originate
either in isolated stars \citep[e.g.][]{Mat00}, or as a result of
close binary systems \citep[e.g.][]{Hol00,Bal02}.  For single stars,
the magnetic fields require a dynamo generated at the interface
between a rapidly rotating stellar core and the more slowly rotating
circumstellar envelope \citep{Bla01}. As the star sheds its outer
layers and leaves the AGB, the core is increasingly exposed and, if
the rotation speed is sufficiently high, the wind is strongly
collimated by magneto-centrifugal forces. The magnetic field lines
remove angular momentum from the core and magnetic braking then slows
down the core rotation. This provides a natural explanation for the
presence of strong collimation in the early post AGB stage, and also
for the observed slow rotation rates of white dwarfs. `Magnetic
explosions' may also occur where energy is released from the wrapping
of magnetic field lines that arise from the magnetised core
\citep{Mat04}. This mechanism may explain the ballistic outflows and
multiple jet axes seen in many PPN.

Although the dynamo models are attractive there is some doubt that the
dynamos can exist for long enough in post-AGB stars to influence the
envelope morphology \citep[e.g.][]{Sok02}. An alternative theory is
that the collimated jets are driven out from the accretion disks of
close binary systems, due to a combination of magnetic and centrifugal
forces.  In this scenario, momentum from the accretion disk is shed
through the production of high-speed jets via an ordered magnetic
field in the disk \citep{Fra04}.  The process is similar to jet
formation from accretion disks around young stellar objects and active
galactic nuclei.  The accretion disk may be formed from the AGB wind
of the primary around a compact secondary companion, or around the
primary from the destruction of the secondary in common envelope
evolution.

The incidence of binarity in post-AGB stars is likely to be $\sim$ 50
per cent \citep[e.g.][]{deR06,Zij06}. From a solar neighbourhood
sample of 164 stars, \citet {Duq91} estimated that $\sim$ 30 per cent
of solar-type stars are isolated, with no companions above 0.01
M$_\odot$. For their sample the median binary period was 180 years,
corresponding to a binary separation of 35 AU.

Binary systems in post-AGB stars are hard to detect as the central
stars are still surrounded by dust. One possible detection method,
through radial velocity variations, is problematic because the orbital 
periods are mostly large enough to require monitoring for at least
several years. In addition, the two central objects must be sufficiently 
close for mass transfer to occur if an accretion disk is to form. 
\citet{Mas99} have estimated that binary separations closer
than about 24 AU are needed for an accretion disk to form.  Given a
median binary separation of 35 AU and a binary fraction of 1/2,
possibly 10--15 per cent of post-AGB stars may be in close binary
systems. Although this is hard to determine with any accuracy, it does
appear that the fraction of evolved stars in close binary systems is
too small to explain the high percentage of PN with aspherical
geometries \citep[e.g.][]{Kwo00}.

For AGB and post-AGB stars with detected maser emission, the incidence
of close binary systems is expected to be small since the masers are
likely to be highly disrupted in such systems. Of the 16 known binary
post-AGB stars in \citet{deR06}, only two objects have OH maser
emission. In searches for OH 1612 MHz maser emission from samples of
symbiotic stars \citep[][sample sizes 16 and 24
respectively]{Nor84,Sea95} there were only two detections, both from
systems thought to be relatively wide binaries. \citet{Sch95} argue
that binaries between 10 and 50 AU will not support OH masers, while
those within 10 AU will not support any maser emission. Thus the
presence of OH emission in particular, will favour single-star or
wide-binary systems. Larger binary systems may show OH masers but
would not be expected to form the accretion disks required to power
the jets.

From the detections of OH masers as well as the high-velocity H$_2$O
emission we consider that while binary systems may be responsible for
the complex structures seen in some PN, the known water-fountain
sources are unlikely to be in close binary systems. The jets may form
from the dynamos associated with the single central stars. This would
imply that dynamos generated in isolated stars can survive for long
enough to support the observed jets. We note that our sources are from 
an OH-selected sample and thus is biased against high-velocity H$_2$O
sources without OH emission. Only one high-velocity H$_2$O source
without OH emission is currently known (IRAS 1934+2134). 

\section{CONCLUSIONS} \label{Sec:water_conclusions}

\begin{itemize}

\item We have searched for 22 GHz H$_2$O maser emission from a
well-defined sample of 85 candidate post-AGB stars. Twenty-one sources
were detected. Of these detections, 12 had `regular' (R) profiles similar to 
those seen in AGB stars. Five had high-velocity (H) profiles, three were
classified as irregular (I) and one has a narrow velocity range near
the stellar velocity (S).

\item Fifteen of the H$_2$O maser detections were from the LI sources. The
OH and H$_2$O maser properties of LI sources are consistent with their
status as massive AGB stars with high mass-loss rates.

\item High-velocity H$_2$O maser emission, with H$_2$O maser features
detected outside of the velocity range of the OH emission, was
detected from five sources. Two of these, v223 (W43A) and v270 (IRAS
19134+2134), have previously been identified. Three sources, d46, b292
and d62 are new discoveries. The source b292 is highly likely to be a
water-fountain source. The source d46 is highly unusual and may be a
more evolved water-fountain source. The status of d62 is less certain
as only one emission feature was detected.

\item For IRAS colours with [12--25] $>$ 1.3 and MSX colours [8--12]
$>$ 0.6, the H$_2$O maser detection rate decreases strongly.  Almost
all sources detected with very red IRAS or MSX colours have high
velocity or irregular outflows.

\item From this survey of 85 sources, we surmise that the H$_2$O
emission seen from AGB stars disappears as a star leaves the AGB. However,
a new source of strong H$_2$O maser emission may appear which
is associated with the emergence of collimated jets from the central star.

\item We consider that the collimated jets seen in water-fountain
sources are likely to be magnetically collimated from the dynamo
action that occurs at the interface of the rapidly rotating stellar core
and more slowly rotating cirucmstellar envelope and argue it is
unlikely that the sources with both OH and H$_2$O masers are in close
binary systems.

\item SiO observations of 11 sources produced three
detections, all from LI stars. One 86 GHz SiO maser source, d47, has
an unusual very wide double-peaked SiO maser profile that may indicate
there is a biconical outflow or torus around this star.
 
\end{itemize}

\section{ACKNOWLEDGMENTS}

We thank the referee of the paper, Mark Claussen, for his constructive
and helpful comments on the manuscript, and Jim Lovell for taking all
of the 22 GHz observations with the Tidbinbilla DSS--43 ratio
telescope. The Deep Space Network DSS--43 antenna is managed by the
Jet Propulsion Laboratory, California Institute of Technology, under a
contract with the National Aeronautics and Space
Administration. Access for radio astronomy is provided through an
agreement between the Australian and US governments and is coordinated
by the Australia Telescope National Facility.

The Mopra radio telescope is part of the Australia Telescope which is
funded by the Commonwealth of Australia for operation as a National
Facility managed by CSIRO.

\end{document}